\documentstyle[12pt]{article}
\begin{document}
\baselineskip = 24pt

\title{ Small angle Bhabha scattering at LEP1. Wide-narrow angular
acceptance.}
\vspace{1.5cm}
\author{N.P.Merenkov}
\date{}
\maketitle
\begin{center}
{\it National Science Centre\footnote{E--mail: kfti@rocket.kharkov.ua}
"Kharkov Institute of Physics and Technology"\\}
{\it Academicheskaya Str.1, 310108, Kharkov, Ukraine}\\
\end{center}
\vspace{1cm}
\begin{abstract} %\small\rm
Analytical method is applied for description  of the small angle Bhabha 
scattering at LEP1. Inclusive event selection for asymmetrical
wide-narrow circular detectors is considered. The QED correction to the
Born cross-section is calculated with leading and next-to-leading accuracy in
the second order of perturbation theory and with leading one -- in the third 
order. All contributions in the second order due to photonic radiative 
corrections
and pair production are calculated starting from essential Feynman diagrams.
The third order correction is computed by means of electron structure function.
Numerical results illustrate the analytical calculsations.
 \end{abstract} \vspace{1.in}
\hspace{2.cm}
PACS \ \ \ 12.15.Lk,\ \ 12.20.-m,\ \ 12.20.Ds,\ \ 13.40.-f

\newpage

\begin{center}
\section {Introduction }
\end{center}

The small angle Bhabha scattering (SABS) process is used to measure the luminosity
of electron-positron colliders. At LEP1 an experimental accuracy on the
luminosity of $\delta\sigma/\sigma <0.1\%$ has been reached [1]. However, to
obtain the total accuracy, a systematic theoretical error must also be added.
The accurate determination the SABS cross-section, therefore, directly affects
some physical values measured at LEP1 experiments\  [2,3]. That is why in recent
years a considerable attention has been devoted to the Bhabha process\  [3-11].
The reached accuracy is, however, still inadequate. According to these
evaluations the theoretical estimates are still incomplete. 

The theoretical calculation of SABS cross-section at LEP1 has to cope with
two somewhat different problems. The first one is the description of an
experimental restrictions used for event selection in terms of final particles
phase space. The second concludes in the writing of matrix element squared
with required accuracy. There are two approaches for theoretical investigation
of SABS at LEP1: the approach based on Monte Carlo calculation [3-5,7] and
semi--analytical one[6,9-11].

The advantage of Monte Carlo method is the possibility to model different
types of detectors and event selection [3]. But at this approach one can not
use in a strightforward way the exact matrix element squared based on essential 
Feynman diagrams because
of infrared divergence. Therefore, some additinal procedures (YFS factor
exponentiation [12], utilization of the electron structure functions [13]) apply
to get rid this problem and to take into account leading contribution in the
higher orders . It needs to be carefully at this point
because of a possibility of the double counting. Any way, up to now the
next-to-leading second order correction remains uncertain, and this is
transparent defect of Monte Carlo approach.

The advantage of analytical method is the possibility to use the exact matrix
element squared. The infrared problem in the frame of this approach is solved
by usual way taking into account virtual, real soft and hard photon emission
as well as pair production in every order of perturbation theory. Any questions
with double counting do not arise at analytical calculations. The defect of
this method is its low mobility relative the change of an experimental
conditions for event selection. Nevertheless, the analytical calculations have
a great importance because allow to check numerous Monte Carlo calculations
for different "ideal" detectors.

Up to now analytical formulae for SABS cross-section at LEP1 are published for
the case of inclusive event selection (IES) when circular symmetrical detectors
record only final electron and positron energies [10,11]. These define the first
and second order corrections to Born cross-section with leading (of the order
$~(\alpha L)^n~$) and next-to-leading (of the order $~\alpha^nL^{n-1}~$)
accuracy as well as third order one with leading only. Just these contributions
will have to be computed to reach required per mille accuracy for SABS
cross-section at LEP1. Note that such accuracy selects only collinear (like
two-jets final-state configuration) and semicollinear (like three-jets one)
kinematics.

In this paper I list full analytical calculation for IES with wide-narrow
angular acceptance. The first and second order corrections are derived with
next-to-leading accuracy starting from Feynman diagrams for two-loops elastic
electron-positron scattering, one-loop single photon emission, two photon
emission and pair production. The third order one is obtained with leading
accuracy by the help of the electron structure function method. The results
for leading second and third order corrections in the case of CES are given
too.

The contents of this paper can be outlined as follows. In Section 2 the
"observable" cross-section $~\sigma_{exp}~$ is introdcued with cuts on angles
and energies taken into account, and the first order correction is obtained.
In Section 3 the second order corrections are investigated. These include
the contributions of the processes of pair (real and virtual) production
considered in Subsection 3.1 and two photons (as well real and virtual)
emission.  In Subsection 3.2 the correction due to one-side two photon 
emission is
considered and in Subsection 3.3 -- due to opposite-side one. The expression 
for the second
order photonic correction is given in leading approximation only, while the
next-to-leading conribution to it is written in Appendix A for both
symmetrical and wide--narrow detectors. The latter does not contain auxiliary
infrared parameter.  In Section 4 the full leading third order correction is
derived using the expansion of electron structure functions. In Section 5 the
numerical results suitable for IES are presented. The correspondence of obtained
results with another semi--analytical ones is dicussed in Conclusion. In 
Appendix B some relations are given which have been used in the process of 
analytical calculations and which will be very useful for numerical ones.
\begin{center}
\section{ First order correcion }
\end{center}

Let us introduce dimentionless quantity
\begin{equation}\label{1}
\Sigma = \frac{1}{4\pi\alpha^2}Q_1^2\sigma_{exp} \ ,
\end{equation}
where $Q_1^2 = \epsilon^2\theta_1^2$ ($\epsilon $ is the beam energy and $
\theta_1$ is the minimal angle of the wide detector). The "experimetally" 
measurable cross section $\sigma_{exp}$ is defined as follows
\begin{equation}\label{2}
\sigma_{exp} = \int dx_1dx_2\Theta d^2q_1^{\bot}d^2q_2^{\bot}\Theta_1^c
\Theta_2^c\frac{d\sigma (e^{+}+e^{-} \rightarrow e^{+}+e^{-}+X)}{dx_1dx_2
d^2q_1^{\bot}d^2q_2^{\bot}} \ ,
\end{equation}
where X is undetected final particles, $x_1 \ (x_2),\ \  q_1^{\bot},\ 
(q_2^{\bot})$
are the energy fraction and the transverse component of the momentum of the
electron (positron) in the final state. Functions $\Theta_{i}^c$ do take
into account angular cuts while function $\Theta$ - cutoff on invariant mass
of detected electron and positron:
$$ \Theta_1^c = \theta(\theta_3 - \theta_{-})\theta(\theta_{-} - \theta_1) \ , \
\Theta_2^c = \theta(\theta_4 - \theta_{+})\theta(\theta_{+} - \theta_2) \ ,\ 
\Theta = \theta (x_1x_2-x_c) \ , $$
\begin{equation}\label{3}
\theta_{-} = \frac{\mid\vec q_1^{\bot}\mid}{x_1\epsilon} \ ,\ \ \
\theta_{+} = \frac{\mid\vec q_2^{\bot}\mid}{x_2\epsilon} \ .
\end{equation}

In the case of symmetrical angular acceptance
$$\theta_2 = \theta_1\ , \  \theta_3 = \theta_4 \ ,\ \  \rho =\frac{\theta_3}
{\theta_1} > 1 \ ,$$
but for wide-narrow one
$$\theta_3 > \theta_4 > \theta_2 > \theta_1 \ ,\ \  \rho_i =\frac{\theta_i}
{\theta_1} > 1 \ .$$
Fof numerical calculation ones usually take
$$\theta_1 = 0.024 \ ,\  \theta_3 =0.058 \ ,\  \theta_2 = 0.024+\frac{0.017}{8}
\ ,\ \theta_4 = 0.058 - \frac{0.017}{8} \ . $$

The first order correction $\Sigma_1$ includes the contributions of virtual
and real soft and hard photon emission processes
\begin{equation}\label{4}
\Sigma_1 = \Sigma_{V+S} + \Sigma^H + \Sigma_H \ .
\end{equation}
The contribution due to virtual and real soft photon (with the energy less
than $ \Delta\epsilon ,\ \Delta\ll1 $ ) may be written as follows ( in this
case $ x_1 = x_2 = 1,\ \  \vec q_1^{\bot} + \vec q_2^{\bot} = 0 $)
\begin{equation}\label{5}
\Sigma_{V+S} = 2\frac{\alpha}{\pi}\int \limits_{\rho_2^2}^{\rho_4^2}
\frac{dz}{z^2}[2(L-1)\ln\Delta +\frac{3}{2}L -2] \ ,\ \   L =
\ln\frac{\epsilon^2 \theta_1^2z}{m^2} \ ,
\end{equation}
where $ z=(\vec q_2^{\bot})^2/Q_1^2$ and \ $m$ is electron mass.

The second term in r.h.s. of Eq.(4) represents the correction due to hard
photon emission by the electron. In this case \begin{equation}\label{6}
X=\gamma(1-x_1, \vec k^{\bot})\ ,\ \  x_2 = 1\ ,\ \  \vec k^{\bot} + \vec q_1^
{\bot} + \vec q_2^{\bot} = 0 \ ,\ \  x_c < x_1 < 1-\Delta \ .
\end{equation}
It can be derived by integration of the differential cross section of single
photon emission over the region
\begin{equation}\label{7}
\rho_2^2 < z < \rho_4^2 \ ,\ \  x^2 < z_1 = \frac{\vec q_1^{\bot 2}}{Q_1^2} <
x^2\rho_3^2 \ ,\ \   -1 < cos\varphi < 1 \ ,
\end{equation}
where $ \varphi $ is the angle between vectors $ \vec q_1^{\bot}$ and $\vec q_2
^{\bot} $, in the same way as it has been done in [10] for symmetrical angular
acceptance. But at this passage I would like to indicate the principle moments
of method used largely to obtain the results of the Section 3  and based on the separate
calculation of the contributions due to collinear kinematics and
semi-collinear one [14].

In collinear kinematics emitted photon moves inside the cone within polar
angles $\theta_{\gamma} < \theta_0 \ll 1$ centred along electron momentum
direction (initial: $\vec k \| \vec p_1 $ or final: $\vec k \| \vec q_1 $). In
semicollinear region photon moves outside this cones. Because such distinction
no longer has physical meaning, the dependence on auxiliary parameter $ \theta_0
$ disappeares in total contribution. This is valid for IES as well as for CES.

Inside collinear kinematics it needs to keep electron mass in differential
cross section
$$ d\sigma = \frac{2\alpha^3s}{\pi^2q^2}\biggl[\frac{1+x^2}{s_1t_1} - \frac{2m^2
}{q^2}\biggl(\frac{1}{s_1^2} + \frac{x^2}{t_1^2}\biggr)\biggr]d\Gamma , $$
\begin{equation}\label{8}
d\Gamma = \frac{d^3q_1d^3q_2d^3k}{\epsilon_1 \omega 2\epsilon}\delta^{(4)}(p_1
+p_2 - k - q_1 - q_2) \ ,
\end{equation}
where $ q=p_1-k-q_1,\  s_1=2(kq_1),\  t_1 = 2(kp_1),\  s=(2p_1p_2)$ \  and \  
$p_1(p_2)$ is 4-momentum
of initial electron (positron). If photon moves inside initial electron cone
$$ s_1 = x(1-x)\epsilon^2\theta_-^2 ,\  t_1 = -m^2(1-x)(1+\eta),\  q^2 = -x^2
\epsilon^2\theta_{-}^2 = - \epsilon^2\theta_{+}^2 \ , $$
\begin{equation}\label{9}
d\Gamma = \frac{m^2}{s}\epsilon^2\pi^2x(1-x)dxd\eta d\theta_{-}^2,\ \   
0 < \eta =
\frac{\theta_{\gamma}^2\epsilon^2}{m^2} < \frac{\theta_0^2\epsilon^2}{m^2} \ ,
\end{equation}
and one can derive after integration relative $\eta$
\begin{equation}\label{10}
\sigma_{\vec k\|\vec p_1} = \frac{2\alpha^3}{Q_1^2}\int \limits_{\rho_2^2}^
{\rho_4^2}\frac{dz}{z^2}\int\limits_{
x_c}^{1-\Delta}dx\biggl[\frac{1+x^2}{1-x}\ln\frac{ \theta_0^2\epsilon^2}{m^2}
- \frac{2x}{1-x}\biggr]\theta(x^2\rho_3^2 - z) \ .  \end{equation} The r.h.s.of
Eq.(10) corresponds to the contribution of narrow strip with the width
$2\sqrt{z}\lambda(1-x)$ centred around line $z = z_1$  in $(z,z_1)$ plane,
where $\lambda = \theta_0/\theta_1$. Really, the condition\  $\theta_{\gamma}
< \theta_0$\  for initial electron cone may be formulated as follows
\begin{equation}\label{11}
\mid\sqrt{z} - \sqrt{z_1}\mid < \lambda(1-x) \ ,\ \   -1 < cos\varphi < -1 +
\frac{\lambda^2(1-x)^2-(\sqrt{z_1}-\sqrt{z})^2}{2\sqrt{z_1z}} \ .
\end{equation}

If photon moves inside final electron cone
$$ s_1 = \frac{1-x}{x}m^2(1+\zeta) \ ,\ t_1 = -(1-x)\epsilon^2\theta_{-}^2 \ , \
q^2 = -\epsilon^2\theta_{-}^2 = -\epsilon^2\theta_{+}^2 \ ,$$
\begin{equation}\label{12}
d\Gamma = \frac{m^2}{s}\epsilon^2\pi^2x(1-x)dxd\zeta\frac{d\theta_{-}^2}{x^2} 
\ , \ \  0 < \zeta < \frac{\theta_{0}^2\epsilon^2x^2}{m^2} \ ,
\end{equation}
and the integration relative $\zeta$ leads to
\begin{equation}\label{13}
\sigma_{\vec k\|\vec q_1} = \frac{2\alpha^3}{Q_1^2}\int\limits_{\rho_2^{^2}}^
{\rho_4^2}\frac{dz}{z^2}\int\limits_{x_c}^{1-\Delta}dx\biggl[\frac{1+x^2}{1-x}
\ln\frac{\theta_0^2\epsilon^2x^2}{m^2} - \frac{2x}{1-x}\biggr] \ .
\end{equation}
The r.h.s. of Eq.(13) corresponds to the contribution of the strip with the
width $ 2\sqrt{z}x^2(1-x)\lambda $ around line $z_1 = x^2z$  in plane
$(z_1,z)$. Really, the condition $\theta_{\gamma}< \theta_{0}$ for final
electron cone may be formulated as $\mid\vec r\mid < \theta_{0}$, where
$\vec r = \vec k/\omega - \vec q_1^{\bot}/\epsilon_1$, and the last
reads as
\begin{equation}\label{14}
\mid\sqrt{z_1} - x\sqrt{z}\mid < x(1-x)\lambda , \ \ -1  < cos\varphi  < -1 +
\frac{\lambda^2x^2(1-x)^2 - (\sqrt{z_1} - x\sqrt{z})^2}{2x\sqrt{zz_1}} \ .
\end{equation}

Having contributions due to collinear regions now it needs to find the
contribution due to semicollinear ones. Supposing $m = 0$ in r.h.s. of Eq.(8)
the differential cross section suitable for this case may be written as
follows \begin{equation}\label{15} d\sigma = \frac{\alpha^3d\varphi
dzdz_1(1+x^2)}{\pi Q_1^2z(z_1-xz)}\biggl[
\frac{1}{z_1+z+2\sqrt{z_1z}cos\varphi} - \frac{x}{z_1+x^2z+2x\sqrt{z_1z}
cos\varphi}\biggr]dx \ .
\end{equation}
When integrating the first term into the brackets in r.h.s. of Eq.(15) one
must use the restriction $\theta_{\gamma} > \theta_{0}$ or
$$\mid\sqrt{z_1} - \sqrt{z}\mid > (1-x)\lambda \ ,\ \  -1  < cos\varphi  < 1 \ 
;$$
\begin{equation}\label{16}
\mid\sqrt{z_1} - \sqrt{z}\mid < (1-x)\lambda \ ,\ \  1  > cos\varphi  > -1 +
\frac{\lambda^2(1-x)^2 - (\sqrt{z_1} - \sqrt{z})^2}{2\sqrt{zz_1}} \ ,
\end{equation}
while for the integration the second one -- the restriction $\mid\vec r\mid >
\theta_{0}$ or
$$\mid\sqrt{z_1} -x\sqrt{z}\mid > x(1-x)\lambda \ ,\ \ -1  < cos\varphi  < 1\ ;
$$
\begin{equation}\label{17}
\mid\sqrt{z_1} - x\sqrt{z}\mid < x(1-x)\lambda \ ,\   1  > cos\varphi  > -1 +
\frac{\lambda^2x^2(1-x)^2 - (\sqrt{z_1} - x\sqrt{z})^2}{2x\sqrt{zz_1}} \ .
\end{equation}

The integration (15) over the region (16) leads to
\begin{equation}\label{18}
\sigma_{a} = \frac{2\alpha^3}{Q_1^2}\int\limits_{\rho_2^2}^
{\rho_4^2}\frac{dz}{z^2}\int\limits_{x_c}^{1-\Delta}\frac{1+x^2}{1-x}dx
\biggl[\biggl(\ln\frac{z}{\lambda^2} + L_2\biggr)\theta_{3}^{(x)} +
L_3\overline \theta_3^{(x)}\biggr].  \end{equation} Analogous, the
integration of r.h.s. of Eq.(15) over the region (17) gives
\begin{equation}\label{19}
\sigma_{b} = \frac{2\alpha^3}{Q_1^2}\int\limits_{\rho_2^2}^
{\rho_4^2}\frac{dz}{z^2}\int\limits_{x_c}^{1-\Delta}\frac{1+x^2}{1-x}dx
\biggl(\ln\frac{z}{x^2\lambda^2} + L_1\biggr) \ .
\end{equation}
The values $L_{i}$ which enter into Eqs.(18) and (19) are defined as follows
$$ L_1 = \ln\left|{x^2(z-1)(\rho_3^2-z)}\over{(x-z)(x\rho_3^2-z)}\right| ,\ \
L_2 = \ln\left|{(z-x^2)(x^2\rho_3^2-z)}\over{x^2(x-z)(x\rho_3^2-z)}\right| ,\
\ L_3 = \ln\left|{(z-x^2)(x\rho_3^2-z)}\over{(x-z)(x^2\rho_3^2-z)}\right| .
$$ Beside this the following notations for $\theta $- functions are used $$
\theta_3^{(x)} = \theta(x^2\rho_3^2-z)\ ,\ \ \overline\theta_3^{(x)} = 1 -
\theta_3^{(x)} = \theta(z-x^2\rho_3^2) \ .$$

Thus, the $\Sigma^{H}$ may be represented as the sum of (10), (13), (18) and
(19) divided by factor $4\pi\alpha^2/Q_1^2$ or
\begin{equation}\label{20}
\Sigma^{H} = \frac{\alpha}{2\pi}\int\limits_{\rho_2^2}^
{\rho_4^2}\frac{dz}{z^2}\int\limits_{x_c}^{1-\Delta}\frac{1+x^2}{1-x}
[(1+\theta_3^{(x)})(L-1) + K(x,z;\rho_3,1)]dx \ ,
\end{equation}
$$ K(x,z;\rho_3,1) = \frac{(1-x)^2}{1+x^2}(1+\theta_3^{(x)}) + L_1 +
\theta_3^{(x)}L_2 + \overline\theta_3^{(x)}L_3 \ . $$
Further I will use the short notations for $\theta$-functions, namely
$$\theta_{i}^{(x)} = \theta(x^2\rho_{i}^2-z) \ ,\  \theta_{i} = \theta(\rho_{i}
^2-z) \ ,\  \overline\theta{i}^{(x)} = 1 - \theta_{i}^{(x)} \ ,\  \ \overline
\theta_{i} = 1 - \theta_{i} \ . $$

One may easy to see that $\Sigma^{H}$ for wide-narrow detectors can be derived
from $\Sigma^{H}$ for symmetrical ones (see[10]) by the change z-integrations
limits
\begin{equation}\label{21}
\int\limits_{1}^{\rho^2}dz \rightarrow   \int\limits_{\rho_2^2}^{\rho_4^2}dz
\end{equation}
and the substitution $\rho_3$ instead of $\rho $ under integral sign.

The third term in r.h.s. of Eq.(4) describes photon emission by the positron. 
It may be derived by full analogy with $\Sigma^{H}$ except restrictions on 
variables $z$ and $z_1$, namely
\begin{equation}\label{22}
1  < z  < \rho_3^2 \ ,\ \ \ \ x^2\rho_2^2 < z_1 < x^2\rho_4^2 \ .
\end{equation}
The contribution of the collinear kinematics ($\vec k \| \vec p_2$ and $\vec k 
\| \vec q_2 $) to single hard photon emission cross section corresponds to the
integration over the regions inside strips with width $ 2\sqrt{z}(1-x)\lambda
$  and $ 2\sqrt{z}x^2(1-x)\lambda $, respectively. It may be written
as follows
$$ \sigma_{\vec k \| \vec p_2, \vec k \| \vec q_2} = \frac{2\alpha^3}{Q_1^2}
\int\limits_{1}^{\rho_3^2}\frac{dz}{z^2}\int\limits_{x_c}^{1-\Delta}
\frac{1+x^2}{1-x}dx\biggl\{\biggl(\ln\frac{\epsilon^2\theta_0^2}{m^2} -
\frac{2x} {1-x}\biggr)\Delta_{42}^{(x)} +$$ \begin{equation}\label{23}
\biggl(\ln\frac{\epsilon^2\theta_0^2x^2}{m^2} - \frac{2x}{1-x}\biggr)
\Delta_{42}\biggr\} \ ,
\end{equation}
where
\begin{equation}\label{24}
\Delta_{42}^{(x)} = \theta_4^{(x)} - \theta_2^{(x)} \ ,\ \ \
\Delta_{42} + \theta_4 - \theta_2 \ .
\end{equation}

The contribution of semi-collinear kinematics may be derived by integration
(15), taking into account the restrictions (16), (17) and (22). The latters
correspond to regions outside narrow strips near $z_1 = z$ and $z_1 = x^2z$,
respectively. The result is
$$ \sigma_{a}+\sigma_{b} = \frac{2\alpha^3}{Q_1^2}\int\limits_{1}^{\rho_3^2}
\frac{dz}{z^2}\int\limits_{x_c}^{1-\Delta}\frac{1+x^2}{1-x}dx\biggl[\ln\frac{z}
{\lambda^2}(\Delta_{42}+\Delta_{42}^{(x)}) + \overline L_2\Delta_{42}^{(x)}
+(\overline L_1 - 2\ln x)\Delta_{42} +$$
\begin{equation}\label{25}
\overline L_3(\overline\theta_4^{(x)} - \theta_2^{(x)}) + \overline L_4(\overline\theta_4 -
\theta_2)\biggr] \ ,
\end{equation}
where $$ \overline L_1 = \ln\left|{(z-\rho_2^2)(\rho_4^2-z)x^2} \over
{(x\rho_4^2-z) (x\rho_2^2-z)}\right|,\ \  \overline L_2
=\ln\left|{(z-x^2\rho_2^2)(x^2\rho_4^2-z)} \over
{x^2(x\rho_4^2-z)(x\rho_2^2-z)}\right| ,$$ \begin{equation}\label{26}
\overline L_3 = \ln\left|{(z-x^2\rho_2^2)(x\rho_4^2-z)} \over {(x^2\rho_4^2-z)
(x\rho_2^2-z)}\right|,\ \ \ \overline L_4
=\ln\left|{(z-\rho_2^2)(x\rho_4^2-z)} \over
{(\rho_4^2-z)(x\rho_2^2-z)}\right| .  \end{equation} The $ \Sigma_{H} $ is
the sum of (23) and (25) divided by $ 4\pi\alpha^2/Q_1^2$:
\begin{equation}\label{27}
\Sigma_{H} = \frac{\alpha}{2\pi}\int\limits_{1}^{\rho_3^2}\frac{dz}{z^2}
\int\limits_{x_c}^{1-\Delta}\frac{1+x^2}{1-x}dx\biggl[(L-1)(\Delta_{42}+
\Delta_{42}^{(x)}) + \widetilde{K}(x,z;\rho_4,\rho_2)\biggr] \ ,
\end{equation}
$$ \widetilde{K} = \frac{(1-x)^2}{1+x^2}(\Delta_{42}+\Delta_{42}^{(x)}) + 
\Delta_{42}\overline L_1 + \Delta_{42}^{(x)}\overline L_2 + (\overline\theta_4^
{(x)} - \theta_2^{(x)})\overline L_3+ (\overline\theta_4 - 
\theta_2)\overline L_4 \ . $$

As one can see the auxiliary parameter $ \theta_0$ disappears in expressions
for $\Sigma^{H} $ and $\Sigma_{H}$, and large logarithm acquires the right
appearence. Thus, the separate investigation of contributions due to collinear
and semi-collinear kinematics simplifies the calculations and gives also the
dipper understanding of underlying physics. The experience of this approach is
very important for the study of CES when it needs to describe events which 
belong to electron cluster (or positron one) in a different way as compared 
with events do not.

The different parts in r.h.s. of Eq.(4) depend on auxiliary infrared paramerter
$\Delta $ but the sum does not. It has the following form:
$$ \Sigma_1 = \frac{\alpha}{2\pi}\biggl\{\int\limits_{1}^{\rho_3^2}\frac{dz}
{z^2}\biggl[-\Delta_{42} + \int\limits_{x_c}^{1}\biggl((L-1)P_1(x)
(\Delta_{42}+\Delta_{42}^{(x)}) + \frac{1+x^2}{1-x}\widetilde
K\biggr)dx\biggr] $$ \begin{equation}\label{28} + \int\limits_{
\rho_2^2}^{\rho_4^2}\frac{dz}{z^2}\biggl[-1 + \int\limits_{
x_c}^{1}\biggl((L-1)P_1(x)(1 + \theta_3^{(x)}) + \frac{1+x^2}
{1-x}K\biggr)dx\biggr]\biggr\} \ ,
\end{equation}
where $$ P_1(x) = \frac{1+x^2}{1-x}\theta(1-x-\Delta) + (2\ln\Delta + \frac{3}
{2})\delta(1-x) \ ,\ \ \Delta \rightarrow 0 \ .$$
In order to make the elimination of $\Delta$ -dependence more transparent one 
can use the following relations:
$$ \int\limits_{x_c}^{1}P_1(x)dx = - \int\limits_{0}^{x_c}\frac{1+x^2}{1-x}dx \ 
, \int\limits_{x_c}^{1}P_1(x)\overline\theta_3^{(x)}dx =  \overline\theta_3^
{(x_c)}\int\limits_{x_c}^{\sqrt{z}/\rho_3}\frac{1+x^2}{1-x}dx \ , $$
\begin{equation}\label{29}
\int\limits_{x_c}^{1}P_1(x)\overline\Delta_{42}^{(x)}dx = \theta_4 \overline
\theta_4^{(x_c)}\int\limits_{x_c}^{\sqrt{z}/\rho_4}\frac{1+x^2}{1-x}dx - \theta
_2 \overline\theta_2^{(x_c)}\int\limits_{x_c}^{\sqrt{z}/\rho_2}\frac{1+x^2}
{1-x}dx \ ,
\end{equation}
where $ \overline\Delta_{42}^{(x)} = \Delta_{42} - \Delta_{42}^{(x)} \ .$

The r.h.s. of Eq.(28) is the full first order QED correction to born SABS cross
section at LEP1 for IES with switched off vacuum polarization. The latter
can be taken into account by insertion the quantity $ [1-\Pi(-zQ_1^2)]^{-2}$
under sign of z-integration (for $\Pi $ see [3] and references therein).
\begin{center}
\section{ Second order correction }
\end{center}

The second order corection contains the contributions due to double photons 
(real and vrtual) emission and pair production. As in symmetrical case one 
needs to distinguish between the situations when additional photons attach only
one fermion line (one-side emission) and two fermion lines (opposite-side 
emission) in corresponding Feynman's diagrams.
\vspace{1cm}
\begin{center}
\subsection{ The contribution of pair production }
\end{center}

Consider at first the contribution of the process of electron-positron pair
production $\Sigma^{pair}$ to the second order correction:
\begin{equation}\label{30}
\Sigma^{pair} = \Sigma^{e^+e^-} + \Sigma_{e^+e^-} \ .
\end{equation}
In order to get rid of the writing some formulae which have the same structure
for both symmetrical and wide-narrow angular acceptance I will often send the
reader to work [11] in which the details of computation are given for
symmetrical case.

The experience of Section 2 allows to write the expression for $\Sigma^{
e^+e^-}$ when created electron-positron pair press to electron momentum
direction, using the result of [11] for  $ \Sigma^{e^+e^-} $ suitable for 
wide--wide angular acceptance. It needs only
to change z-integration limits; $(\rho^2, 1) \rightarrow (\rho_4^2, \rho_2^2)
$ and substitute $ \rho_3 $ instead of $ \rho $ everywhere under integral sign.
The result may be written as follows:
$$ \Sigma^{e^+e^-} = \frac{\alpha^2}{4\pi^2}\int\limits_{\rho_2^2}^{\rho_4^2}
\frac{dz}{z^2}L\biggl\{L\biggl(1 + \frac{4}{3}\ln(1-x_c) - \frac{2}{3}\int
\limits_{x_c}^{1}\frac{dx}{1-x}\overline\theta_3^{(x)}\biggr) - \frac{17}{3} -
\frac{8}{3}\zeta_2 - $$
$$- \frac{40}{9}\ln(1-x_c) +
\frac{8}{3}\ln^2(1-x_c) + \int\limits_{\ \ \
x_c}^{1}\frac{dx}{1-x}\overline\theta_3^ {(x)}\biggl(\frac{20}{9} -
\frac{8}{3}\ln(1-x)\biggr) + $$ 
\begin{equation}\label{31}
+ \int\limits_{x_c}^{1}\biggl[L\overline R(x)(1+\theta_3^{(x)}) +
\theta_3^{(x)}C_1(x,z;\rho_3) + C_2(x) + d_2(x,z;\rho_3)\biggr]dx\biggr\} \ ,
\end{equation}
$$ \overline R(x) = (1+x)(\ln x - \frac{1}{3}) + \frac{1-x}{6x}(4 + 7x + 4x^2)
\ , $$ 
$$C_1(x,z;\rho_3) = - \frac{113}{9} + \frac{142}{9}x -
\frac{2}{3}x^2 - \frac {4}{3x} - \frac{4}{3}(1+x)\ln(1-x) +
\frac{2(1+x^2)}{3(1-x)}\biggl[2\ln\left| {x^2\rho_3^2 - z} \over {x\rho_3^2 -
z}\right| - $$ $$- 3L_{i2}(1-x)\biggr] +
(8x^2+3x-9-\frac{8}{x}-\frac{7}{1-x})\ln x + \frac{2(5x^2-6)}{1-x}\ln^2x +
R(x)\ln\frac{(x^2\rho_3^2-z)^2}{\rho_3^4} \ ,$$
$$C_2(x) = - \frac{122}{9} + \frac{133}{9}x + \frac{4}{3}x^2 + \frac
{2}{3x} - \frac{4}{3}(1+x)\ln(1-x) + \frac{2(1+x^2)}{(1-x)}L_{i2}(1-x) + $$
$$+\frac{1}{3}(-8x^2-32x-20+\frac{8}{x}+\frac{13}{1-x})\ln x + 3(1+x)\ln^2x,
\ \ R(x) =  2\overline R(x) + \frac{2}{3}(1+x) \ , $$
\begin{equation}\label{32}
d_2(x,z;\rho_3) = \frac{2(1+x^2)}{3(1-x)}\ln\left|{(z-x^2)(\rho_3^2-z)(z-1)} 
\over{(z-x)^2(x^2\rho_3^2-z)}\right| + 
+ R(x)\ln\left|{(z-x^2)(\rho_3^2-z)(z-1)}\over {x^2\rho_3^2-z}\right| \ .
\end{equation}

The r.h.s. of Eq.(31) does not contain infrared auxiliary parameter because it
includes the contributions due to real and virtual pair production. The
contribution of hard pair takes into account both, collinear and semi-collinear
kinematics, and this ensures the next-to-leading accuracy.

If created elctron-positron pair is emitted along of the positron momentum 
direction the corresponding expression requires more modifications. The source 
of such modifications is the semi-collinear kinematics as we saw in Section 2 
for the single photon emission.

The strightforward calculation shows that for contribution of the semi-collinear
region $\vec p_+ \| \vec p_- $ (I use here notation $ \vec p_{\pm} $ for 3 -
momentum of created positron (electron)) one has to write into formula (28) of
[11]
$$(\Delta_{42} + \Delta_{42}^{(x)})\ln\frac{z}{\lambda^2} +
\Delta_{42}\ln\left |{(z-\rho_2^2)(\rho_4^2-z)} \over
{(z-x\rho_2^2)(x\rho_4^2-z)}\right| +
\Delta_{42}^{(x)}\ln\left|{(z-x^2\rho_2^2)(x^2\rho_4^2-z)}\over {x^2(z-x\rho_
2^2)(x\rho_4^2-z)}\right| + $$
\begin{equation}\label{33}
(\overline\theta_4-\theta_2)\ln\left|{(z-\rho_2^2)(x\rho_4^2-z)} \over
{(z-x\rho_2^ 2)(\rho_4^2-z)}\right|
+(\overline\theta_4^{(x)}-\theta_2^{(x)})\ln\left|{(z-x^2
\rho_2^2)(x\rho_4^2-z)} \over {(z-x\rho_2^2)(z-x^2\rho_4^2)}\right|
\end{equation}
instead of expression in curle brackets and change the upper limit of 
z-integration: $\rho \rightarrow \rho_3 \ . $

For the contribution of semi-collinear region $\vec p_+ \| \vec q_1$ the
correspnding expression is (see Eq.(33) in [11])
\begin{equation}\label{34}
\Delta_{42}\biggl(\ln\frac{z}{\lambda^2} + \ln\left|{(z-\rho_2^2)(\rho_4^2-z)}
\over {x_2^2\rho_2^2\rho_4^2}\right|\biggr) +
(\overline\theta_4-\theta_2)\ln\left| {\rho_4^2(z-\rho_2^2)} \over
{\rho_2^2(z-\rho_4^2)}\right| , \end{equation} and for semi-collinear region
$ \vec p_- \| \vec p_1 $ (see Eq.(38) in [11]) \begin{equation}\label{35}
\Delta_{42}^{(x)}\biggl(\ln\frac{z}{\lambda^2} + \ln\left|{(z-x^2\rho_2^2)
(x^2\rho_4^2-z)}\over {x_1^2x^4\rho_2^2\rho_4^2}\right|\biggr) +
(\overline\theta_4^{(x)}-\theta_2^{(x)})\ln\left|
{\rho_4^2(z-x^2\rho_2^2)} \over {\rho_2^2(z-x^2\rho_4^2)}\right| .
\end{equation}
For the symmetrical wide--wide angular acceptance $\rho_3 = \ \rho_4 = \ \rho \
, \ \ \rho_2 = 1 \ ,$ and
\begin{equation}\label{36}
\Delta_{42} \rightarrow \theta(\rho^2-z)\theta(z-1)\ ,\  \Delta_{42}^{(x)}
\rightarrow \theta(x^2\rho^2-z)\ ,\  \overline\theta_4^{(x)} \rightarrow 
\theta(z-x^2\rho^2)\ ,\ \overline\theta_4\ ,\ \theta_2,\ \theta_2^{(x)} 
\rightarrow 0 \ ,
\end{equation}
and (33), (34), (35) reduce to corresponding expressions derived in [11]\ .

The modification of the contributions due to virtual, real soft and hard
collinear pair production includes the change of z-integral upper limit :
$\rho \rightarrow \rho_3 $ and trivial change of $\theta -$functions under
integral sign, namely: \  $ \theta(x^2\rho^2-z) \rightarrow \Delta_{42}^{(x)},
\ \ 1 \rightarrow \Delta_{42}.$ The sum of all contributions has the
following form:  $$ \Sigma_{e^+e^-} = \frac{\alpha^2}{4\pi^2}\int\limits_{
1}^{\rho_3^2} \frac{dz}{z^2}L\biggl\{L\biggl[\Delta_{42}(1 +
\frac{4}{3}\ln(1-x_c)) - \frac{2}{3}\int\limits_{
x_c}^{1}\frac{dx}{1-x}\overline\Delta_{42}^{(x)}\biggr] + \Delta_{42}\biggl(-
\frac{17}{3} -\frac{8}{3}\zeta_2 - $$ 
$$- \frac{40}{9}\ln(1-x_c) + \frac{8}{3}\ln^2(1-x_c) \biggr)+ \int\limits_{
x_c}^{1}\frac{dx}{1-x}\overline \Delta_{42}^{(x)}\biggl(\frac{20}{9} -
\frac{8}{3}\ln(1-x)\biggr) + \int\limits_{x_c}^{1}\biggl[L\overline
R(x)(\Delta_{42}+\Delta_{42}^{(x)}) + $$ 
$$+\Delta_{42}^{(x)}C_1(x,z;\rho_2) + \Delta_{42}(C_2(x) + \overline
d_2(x,z;\rho_2) ) +
(\overline\theta_4^{(x)}-\theta_4^{(x)})\biggl(\frac{2(1+x^2)}{3(1-x)}\ln\left|{(x^2\rho_2
^2-z)(x\rho_4^2-z)} \over {(x^2\rho_4^2-z)(x\rho_2^2-z)}\right| + $$
$$+ R(x)\ln\left|{(x^2\rho_2^2-z)\rho_4^2} \over {(x^2\rho_4^2-z)\rho_2^2}
\right|\biggr)+ (\overline\theta_4 -
\theta_4)\biggl(\frac{2(1+x^2)}{3(1-x)}\ln \left|{(x\rho_4^2-z)(z-\rho_2^2)}
\over {(x\rho_2^2-z)(z-\rho_4^2)}\right| +$$ \vspace{0.2cm}
$$+ R(x)\ln\left|{(\rho_2^2-z)\rho_4^2} \over {(\rho_4^2-z)\rho_2^2}\right|
\biggr)\biggr]\biggr\} \ , $$
\begin{equation}\label{37}
\overline d_2(x,z;\rho_2) = \frac{2(1+x^2)}{3(1-x)}\ln\frac{(z-\rho_2^2)^2}
{(z-x\rho_2^2)^2} + 2R(x)\ln\frac{z-\rho_2^2}{\rho_2^2} \ .
\end{equation}
By the help of (36) one can verify that r.h.s. of Eq.(36) goes over in
corresponding expression for symmetrical angular acceptance.
\begin{center}
\subsection{ The contribution of one-side double photon emission }
\end{center}

In this Section I give the analytical expressions for all contributions into
the second order correction which appear due to one-side two photon (real and
virtual) emission. The master formula which does not contain infrared auxiliary
parameter $\Delta$ is written only for leading approximation, and
next-to-leading contribution to it is given in Apendix A.

As before it needs to differ the radiation along electron and positron momentum
directions
$$\Sigma_2 = \Sigma^{\gamma\gamma} + \Sigma_{\gamma\gamma} \ ,\ \ \Sigma^{\gamma
\gamma} = \Sigma^{(S+V)^2} + \Sigma^{(S+V)H} + \Sigma^{HH}, $$
\begin{equation}\label{38}
\Sigma_{\gamma\gamma} = \Sigma_{(S+V)^2} + \Sigma_{(S+V)H} + \Sigma_{HH}\ .
\end{equation}

The contribution of virtual and real soft photon is the same for both the 
electron and the positron emission
$$ \Sigma_{(S+V)^2} = \Sigma^{(S+V)^2} = \frac{\alpha^2}{\pi^2}\int\limits_
{\rho_2^2}^{\rho_4^2}\frac{dz}{z^2}L\biggl[L(2\ln^2\Delta +
3\ln\Delta +\frac{9} {8})- $$  \begin{equation}\label{39}
4\ln^2\Delta - 7\ln\Delta + 3\zeta_3 - \frac{3}{2}\zeta_2 -
\frac{45}{16}\biggr] \ .  \end{equation}

Virtual and real soft photon correction to single hard photon emission already
differs for photon moving along the electron momentum direction and the 
positron one. In the first case corresponding contribution may be derived by 
the help of result for symmetrical detector (see[10], formula(50)) using the
substitutions $(\rho_4^2,\ \rho_2^2)$ instead of $ (\rho^2,\ 1) $ for
z-integration limits and $ \rho_3 $ instead of $\rho $ under integral sign.
Therefore, $$ \Sigma^{(S+V)H} = \frac{\alpha^2}{2\pi^2}\int\limits_{
\rho_2^2}^{\rho_4^2} \frac{dz}{z^2}L\int\limits_{
x_c}^{1-\Delta}\frac{1+x^2}{1-x}dx\biggl\{(2\ln \Delta - \ln x
+\frac{3}{2})\biggl[K(x,z;\rho_3,1) + $$ 
$$+ (L-1)(1+\theta_3^{(x)})\biggr] + \frac{1}{2}\ln^2x -
\frac{(1-x)^2}{2(1+x^2)} + (1+\theta_3^{(x)})(-2+\ln x-2\ln\Delta) +
\overline\theta_3^{(x)}\biggl[\frac{1}{2} L\ln x + $$
\begin{equation}\label{40}
+ 2\ln\Delta \ln x - \ln x\ln(1-x) - \ln^2x - L_{i2}(1-x) -
\frac{x(1-x)+4x\ln x}{2(1+x^2)} \biggr]\biggr\} \ .  \end{equation} In order to
obtain the expression for $\Sigma_{(S+V)H} $ it needs to change in r.h.s. of
Eq.(39): $$ \mbox{ i) limits of z-integration:}\; (\rho_4^2,\ \rho_2^2) \rightarrow
(\rho_3^2,\ 1) \ ,$$ 
\begin{equation}\label{41} ii) \ K(x,z:\rho_3,1) \rightarrow
\widetilde{K}(x,z:\rho_4,\rho_2)\ ;\ \theta_3^{(x)} \rightarrow
\Delta_{42}^{(x)}\ ,\ \overline\theta_3^{(x)} \rightarrow \overline\Delta_{42}^{(x)},\ \
1 \rightarrow \Delta_{42}\ .
\end{equation}
The contribution of two hard photons emitted along electron momentum directon
may be obtained in the same way as $ \Sigma^{(S+V)H} $, using the known result
for symmetrical detectors (see [10], Eq.(54)), namely:
\begin{equation}\label{42}
\Sigma^{HH} = \frac{\alpha^2}{4\pi^2}\int\limits_{\rho_2^2}^{\rho_4^2}\frac{dz}
{z^2}L\int\limits_{x_c}^{1-2\Delta}dx\int\limits_{\Delta}^{1-x-\Delta}dx_1
\frac{I^{HH}}{x_1(1-x-x_1)(1-x_1)^2} \ ,
\end{equation}
$$ I^{HH} = \overline A\theta_3^{(x)} + \overline B + \overline C\theta_3^{(1-x_1)} ,  $$
$$ \overline A = \gamma\beta\biggl(\frac{L}{2} +
\ln\frac{(x^2\rho_3^2-z)^2}{x^2 (x(1-x_1)\rho_3^2-z)^2}\biggr)+ \zeta
\ln\frac{(1-x_1)^2(1-x-x_1)}{xx_1} + \gamma_A \ , $$  $$
\overline B = \gamma\beta\biggl(\frac{L}{2} +
\ln\left|{x^2(z-1)(\rho_3^2-z)(z-x^2) (z-(1-x_1)^2)^2(\rho_3^2x(1-x_1)-z)^2}
\over {(\rho_3^2(1-x_1)^2-z)^2
(z-(1-x_1))^2(z-x(1-x_1))^2(\rho_3^2x^2-z)}\right|\biggr) + $$
$$+ \zeta \ln\frac{(1-x_1)^2x_1}{x(1-x-x_1)} + \delta_B \ , $$
\begin{equation}\label{43}
\overline C = \gamma\beta\biggl(L + 2\ln\left|{x(\rho_3^2(1-x_1)^2-z)^2} \over
{(1-x_1)^2(\rho_3^2x(1-x_1)-z)(\rho_3^2(1-x_1)-z)}\right|\biggr) -
2(1-x_1)\beta-2x(1-x_1)\gamma \ ,
\end{equation}
where $$ \gamma = 1 + (1-x_1)^2,\ \ \beta = x^2 + (1-x_1)^2, \ \ \zeta =
x^2 + (1-x_1)^4,$$
$$ \gamma_A = xx_1(1-x-x_1)-x_1^2(1-x-x_1)^2-2(1-x_1)\beta,\ \
\delta_B = xx_1(1-x-x_1)-x_1^2(1-x-x_1)^2-2x(1-x_1)\gamma .$$
Unfortunately, it is impossible to give such simple prescription as (41) in
order to obtain $\Sigma_{HH}$ from Eqs.(42) and (43). In the case of radiation
two hard photons along the positron momentum direction an additional detailed
consideration of semi-collinear kinematics is required. All essential moments
of such consideration shown in Section 2, and reader can make all calculations
by the help of formulae given in Appendix B of ref.[10]. Here I give final
result \begin{equation}\label{44} \Sigma_{HH} =
\frac{\alpha^2}{4\pi^2}\int\limits_{1}^{\rho_3^2}\frac{dz}
{z^2}L\int\limits_{x_c}^{1-2\Delta}dx\int\limits_{\Delta}^{1-x-\Delta}dx_1
\frac{I_{HH}}{x_1(1-x-x_1)(1-x_1)^2} \ ,
\end{equation}
$$ I_{HH} = \widetilde A\Delta_{42}^{(x)} + \widetilde C\Delta_{42}^{(1-x_1)} +
\widetilde B\Delta_{42} + (\overline\theta_4^{(x)}-\theta_2^{(x)}){\it a} +
(\overline\theta_4^{(1-x_1)}-\theta_2^{(1-x_1)}){\it c} + (\overline\theta_4-\theta_2)
{\it b} \ ,$$
$${\it a} = \gamma\beta \ln\left|{(\rho_4^2x(1-x_1)-z)(\rho_2^2x^2-z)} \over
{(\rho_2^2x(1-x_1)-z)(\rho_4^2x^2-z)}\right|\ ,\ \ \
{\it b} = \gamma\beta \ln\left|{(\rho_4^2(1-x_1)-z)(\rho_2^2-z)} \over
{(\rho_2^2(1-x_1)-z)(\rho_4^2-z)}\right| ,$$
$${\it c} = \gamma\beta \ln\left|{(\rho_4^2x(1-x_1)-z)(\rho_2^2(1-x_1)^2-z)^2
(\rho_4^2(1-x_1)-z)} \over {(\rho_2^2x(1-x_1)-z)(\rho_4^2(1-x_1)^2-z)^2
(\rho_2^2(1-x_1)-z)}\right| , $$
$$ \widetilde A = \gamma\beta\biggl(\frac{L}{2} + \ln\left|{(\rho_4^2x^2-z)
(\rho_2^2x^2-z)} \over {x^2(\rho_4^2x(1-x_1)-z)(\rho_2^2x(1-x_1)-z)}\right|
\biggr) + \zeta \ln\frac{(1-x_1)^2(1-x-x_1)}{xx_1} + \gamma_A \ , $$
$$ \widetilde B = \gamma\beta\biggl(\frac{L}{2} + \ln\left|{x^2(\rho_4^2-z)
(\rho_2^2-z)} \over {(\rho_4^2(1-x_1)-z)(\rho_2^2(1-x_1)-z)}\right|
\biggr) + \zeta \ln\frac{(1-x_1)^2x_1}{x(1-x-x_1)} + \delta_B \ , $$
$$ \widetilde C = \gamma\beta\biggl(L + \ln\left|{x^2(\rho_4^2(1-x_1)^2-z)^2
(\rho_2^2(1-x_1)^2-z)^2} \over {(1-x_1)^4(\rho_4^2x(1-x_1)-z)
(\rho_2^2x(1-x_1)-z)(\rho_4^2(1-x_1)-z)(\rho_2^2(1-x_1)-z)}\right|\biggr)$$
$$-2(1-x_1)(\beta + x\gamma) \ . $$

As one can see the separate contributions in r.h.s. of Eq.(38) depend on
infrared auxiliary parameter $\Delta $ but $\Sigma^{\gamma\gamma}$ and
$\Sigma_{\gamma\gamma} $ do not. In order to eliminate $\Delta $-dependence
analytically it needs to apply a lot efforts. Below I give leading terms and
for next-to-leading ones see Appendix A.
\begin{equation}\label{45}
\Sigma^{\gamma\gamma L} = \frac{\alpha^2}{4\pi^2}\int\limits_{\rho_2^2}^
{\rho_4^2}\frac{dz}{z^2}L^2\int\limits_{x_c}^{1}dx\biggl[\frac{1}{2}(1+\theta
_3^{(x)})P_2(x) + \int\limits_{x}^{1}\frac{dt}{t}P_1(t)P_1\biggl(\frac{x}{t}
\biggr)\theta_3^{(t)}\biggr] ,
\end{equation}
\begin{equation}\label{46}
\Sigma_{\gamma\gamma}^L = \frac{\alpha^2}{4\pi^2}\int\limits_{1}^{\rho_3^2}
\frac{dz}{z^2}L^2\int\limits_{x_c}^{1}dx\biggl[\frac{1}{2}(\Delta_{42}+
\Delta_{42}^{(x)})P_2(x) + \int\limits_{x}^{1}\frac{dt}{t}P_1(t)P_1
\biggl(\frac{x}{t}\biggr)\Delta_{42}^{(t)}\biggr] ,
\end{equation}
where
$$ P_2(x) = P_1\otimes P_1 = \int\limits_{\ \ \ x}^{1}\frac{dt}{t}P_1(t)P_1
\biggl(\frac{x}{t}\biggr) = \lim_{\Delta \to 0} \biggl\{\biggl[(2\ln\Delta +
\frac{3}{2})^2 - 4\zeta_2\biggr]\delta(1-x) + $$
\begin{equation}\label{47}
+ 2\biggl[\frac{1+x^2}{1-x}(2\ln(1-x)-\ln x+\frac{3}{2}) +
\frac{1}{2}(1+x)\ln x -1 +x\biggr]\theta(1-x-\Delta)\biggr\}, \end{equation}
$$\int\limits_{\ \ \ 0}^{1}P_2(x)dx = 0 \ . $$ The expressions (45) and (46)
are not convenient for numerical calculations.  The suitable ones may be
written as follows \begin{equation}\label{48} \Sigma^{\gamma\gamma L} =
\frac{\alpha^2}{4\pi^2}\biggl\{-2\int\limits_{\rho_2^2}
^{\rho_4^2}\frac{dz}{z^2}L^2\int\limits_{0}^{x_c}P_2(x)dx - \int\limits_
{m_{23}}^{\rho_4^2}\frac{dz}{z^2}L^2\int\limits_{x_c}^
{\sqrt{z}/\rho_3}\biggl[P_1(x)g\biggl(\frac{x_c}{x}\biggr) + \frac{1}{2}P_2(x)
\biggr]dx\biggr\}\ ,
\end{equation}
\begin{equation}\label{49}
\Sigma_{\gamma\gamma}^L = \frac{\alpha^2}{4\pi^2}\biggl\{-2\int\limits_
{\rho_2^2}^{\rho_4^2}\frac{dz}{z^2}L^2\int\limits_{0}^{x_c}P_2(x)dx -
\int\limits_{m_{14}}^{\rho_4^2}\frac{dz}{z^2}L^2\int\limits_{x_c}^
{\sqrt{z}/\rho_4}\biggl[P_1(x)g\biggl(\frac{x_c}{x}\biggr) + \frac{1}{2}P_2(x)
\biggr]dx\biggr\} +
\end{equation}
$$+ \int\limits_{m_{12}}^{\rho_2^2}\frac{dz}{z^2}L^2\int\limits_{x_c}^
{\sqrt{z}/\rho_2}\biggl[P_1(x)g\biggl(\frac{x_c}{x}\biggr) + \frac{1}{2}P_2(x)
\biggr]dx\biggr\} ,$$
where $$ g(y) = y +\frac{y^2}{2} + 2\ln(1-y)\ ,\ \ m_{23} = max(\rho_2^2\ ,\
x_c^2\rho_3^2)\ , $$
$$ m_{14} = max(1,\ x_c^2\rho_4^2) \ ,\ \ m_{12} = max(1,\ x_c^2\rho_2^2)\ . $$
The last two formulae can be derived by means the relations given in Appendix
B. The integration relative $x$-variable in Eqs.(45) and (46) may be performed
by the help of the following formulae
\begin{equation}\label{50}
\int\limits_{\ }^{x}P_2(y)dy=F_2(x)\ ,\quad \int\limits_{\ }^{x}P_1(y)g\biggl(
\frac{x_c}{y}\biggr)dy = F_g(x)\ , \ \
\int\limits_{\ }^{x}P_1(y)dy = -g(x)\ , \ \ x < 1\ ,
\end{equation}
\begin{equation}\label{51}
F_2(x) = -2x - \frac{x^2}{4} + (x+\frac{x^2}{2})\ln\frac{x^3}{(1-x)^4} +
4\ln(1-x)\ln\frac{x}{1-x} + 4L_{i2}(x) \ ,
\end{equation}
$$F_g(x) = -\frac{x_c^2}{2x} + (2x+x^2)\ln x +
(x_c+\frac{x_c^2}{2})\ln\frac{x} {(1-x)^2} +
(2x_c+\frac{x_c^2}{2}-2x-\frac{x^2}{2})\ln(x-x_c) +$$
\begin{equation}\label{52}
+4L_{i2}(x) + 4L_{i2}\biggl(\frac{1-x}{1-x_c}\biggr)\ , \quad x_c < x < 1 \ .
\end{equation}
Therefore, the second order leading contribution to SABS cross section at
LEP1 can be expressed through integral relative z-variable only.

It is useful to note also that for CES the leading contributions in all orders
of perturbation theory take into account the emission of photons in initial
state only. Thus, the corresponding correction due to one-side two photon
(real and virtual) emission will be read in this case as follows:
\begin{equation}\label{53}
\Sigma_{CES}^{\gamma\gamma L} = - \frac{1}{8}\biggl(\frac{\alpha}{\pi}\biggr)
^2\int\limits_{\ \ \ \rho_2^2}^{\rho_4^2}\frac{dz}{z^2}L^2\biggl\{F_2(x_c) +
\biggl[F_2\biggl(\frac{\sqrt{z}}{\rho_3}\biggr) - F_2(x_c)\biggr]\overline\theta_3
^{(x_c)}\biggr\},
\end{equation}
\begin{equation}\label{54}
\Sigma_{\gamma\gamma \ CES}^L = - \frac{1}{8}\biggl(\frac{\alpha}{\pi}\biggr)^2
\biggl\{\int\limits_{\ \ \ \rho_2^2}^{\rho_4^2}\frac{dz}{z^2}L^2F_2(x_c) +
\int\limits_{\ \ \ 1}^{\rho_4^2}\frac{dz}{z^2}L^2\biggl[F_2\biggl(\frac
{\sqrt{z}}{\rho_4}\biggr) - F_2(x_c)\biggr]\overline\theta_4^{(x_c)} -
\end{equation}
$$- \int\limits_{\ \ \ 1}^{\rho_2^2}\frac{dz}{z^2}L^2\biggl[F_2\biggl(\frac
{\sqrt{z}}{\rho_2}\biggr) - F_2(x_c)\biggr]\overline\theta_2^{(x_c)} \biggr\} .$$
\begin{center}
\subsection{ Second order correction due to opposite-side photon
emission } \end{center}

In this Section I calculate analytically the expression for
\begin{equation}\label{55}
\Sigma_{\gamma}^{\gamma} = \Sigma_{S+V}^{S+V} + \Sigma_{S+V}^{H} +
\Sigma_{H}^{S+V} + \Sigma_{H}^{H} .
\end{equation}
The quantity $ \Sigma_{\gamma}^{\gamma} $ does not depend on infrared auxiliary
parameter $\Delta$ because it contains all contributions due to virtual, real
soft and hard photon emission.

The first term in r.h.s. of Eq.(55) takes into account only "oposite-side"
virtual and real soft photon corrections
\begin{equation}\label{56}
\Sigma_{S+V}^{S+V} = \frac{\alpha^2}{\pi^2}\int\limits_{\rho_2^2}^{\rho_4
^2}\frac{dz}{z^2}L\biggl[L(4\ln^2\Delta + 6\ln\Delta +\frac{9}{4}) - 6 - 14
\ln\Delta - 8\ln^2\Delta\biggr] \ .
\end{equation}

The contribution of one-loop virtual and real soft photon corrections to hard
single photon emission may be written as follows
\begin{equation}\label{57}
\Sigma_{S+V}^{H} = \frac{\alpha^2}{2\pi^2}\int\limits_{\rho_2^2}^
{\rho_4^2}\frac{dz}{z^2}\biggl[2(L-1)\ln\Delta + \frac{3}{2}L - 2\biggr]
\int\limits_ {x_c}^{1-\Delta}\frac{1+x^2}{1-x}\biggl[(1 + \theta_3^
{(x)})(L-1) +K(x,z;\rho_3,1)\biggr]\ ,
\end{equation}
\begin{equation}\label{58}
\Sigma_{H}^{S+V} = \frac{\alpha^2}{2\pi^2}\int\limits_{\ \ \ 1}^{\rho_3^2}
\frac{dz}{z^2}\biggl[2(L-1)\ln\Delta + \frac{3}{2}L - 2\biggr]\int\limits_
{\ \ \ x_c}^{1-\Delta}\frac{1+x^2}{1-x}\biggl[(\Delta_{42} + \Delta_{42}^{(x)})(L-1) +
\widetilde K(x,z;\rho_4,\rho_2)\biggr]dx.
\end{equation}

In order to find the contribution of two opposite-side hard photon emission
into $~\Sigma_{\gamma}^{\gamma}~$ it is convenient to use the factorization
theorem for differential cross-sections of two-jets processes in QCD [16]. It
reads as:
\begin{equation}\label{59}
\Sigma_{H}^{H} = \frac{\alpha^2}{4\pi^2}\int\limits_{0}^{\infty}
\frac{dz}{z^2}\int\limits_{x_c}^{1-\Delta}dx_1\int\limits_{
{x_c}/{x_1}}^{1-\Delta}dx_2\frac{1+x_1^2}{1-x_1}\frac{1+x_2^2}{1-x_2}
\Phi(x_1,z,;\rho_3,1)\Phi(x_2,z;\rho_4,\rho_2)\ ,
\end{equation}
$$\Phi(x,z,;\rho_3,1) = (\Delta_{31} + \Delta_{31}^{(x)})(L-1) + \frac{(1-x)^2}
{1+x^2}(\Delta_{31} + \Delta_{31}^{(x)}) + \Delta_{31}L_1 + \Delta_{31}^{(x)}
L_2 \ ,$$
\begin{equation}\label{60}
(\overline\theta_3^{(x)} - \theta_1^{(x)})L_3 + (\overline\theta_3 - \theta_1)
\ln\left|{(x\rho_3^2-z)(z-1)} \over {(z-x)(\rho_3^2-z)}\right| ,
\end{equation}
\begin{equation}\label{61}
\Phi(x,z,;\rho_4,\rho_2) = (\Delta_{42} + \Delta_{42}^{(x)})(L-1) +
\widetilde K(x,z;\rho_4,\rho_2)\ ,
\end{equation}
$$\Delta_{31} = \theta_3 - \theta_1\ , \quad \Delta_{31}^{(x)} = \theta_3^{(x)}
 - \theta_1^{(x)}, \quad \theta_1 = \theta(1-z)\ , \quad \theta_1^{(x)} =
\theta(x^2-z) \ .$$

The $\Delta$-dependence of separate terms in r.h.s. of Eq.(55) can be eliminated
analytically in the whole sum. The leading contribution is expressed in terms
of electron structure functions as follows
\begin{equation}\label{62}
\Sigma_{\gamma}^{\gamma L} = \frac{\alpha^2}{4\pi^2}\int\limits_{0}^
{\infty}\frac{dz}{z^2}L^2\int\limits_{x_c}^{1}dx_1\int\limits_
{{x_c}/{x_1}}^{1}dx_2P_1(x_1)P_1(x_2)(\Delta_{31} + \Delta_{31}^{(x_1)})
(\Delta_{42} + \Delta_{42}^{(x_2)}) \ .
\end{equation}
The next-to-leading contribution to $\Sigma_{\gamma}^{\gamma}$ is given in
Appendix A.

The form of $\Sigma_{\gamma}^{\gamma}$ suitable for numerical counting may be
written in terms of functions $ F_2(x)$ and $F_g(x) $ in the same manner as it
was done at the end of Subsection 3.2
$$\Sigma_{\gamma}^{\gamma L} = \frac{\alpha^2}{4\pi^2}\biggl\{ -\int\limits_
{\rho_2^2}^{\rho_4^2}\frac{dz}{z^2}L^2\biggl[4(1)F_2(x_c) + 2(1)\biggl(
F_g\biggl(\frac{\sqrt{z}}{\rho_3}\biggr) - F_g(x_c)\biggr)\overline\theta_3^
{(x_c)} - $$
$$-\int\limits_{1}^{\rho_4^2}\frac{dz}{z^2}L^2 2(1)\biggl(F_g\biggl(
\frac{\sqrt{z}}{\rho_4}\biggr) - F_g(x_c)\biggr)\overline\theta_4^{(x_c)} +
\int\limits_{1}^{\rho_2^2}\frac{dz}{z^2}L^2 2(1)\biggl(F_g\biggl(
\frac{\sqrt{z}}{\rho_2}\biggr) - F_g(x_c)\biggr)\overline\theta_2^{(x_c)} + $$
$$ + \int\limits_{x_c\rho_3\rho_4}^{\rho_4^2}\frac{dz}{z^2}L^2\biggl[
F_g\biggl(\frac{\sqrt{z}}{\rho_4}\biggr) - F_g\biggl(\frac{x_c\rho_3}
{\sqrt{z}}\biggr) + g\biggl(\frac{\sqrt{z}}{\rho_3}\biggr)\biggl(g\biggl
(\frac{\sqrt{z}}{\rho_4}\biggr) - g\biggl(\frac{x_c\rho_3}{\sqrt{z}}\biggr)
\biggr)\biggr] + $$
$$+ \int\limits_{x_c\rho_2}^{1}\frac{dz}{z^2}L^2\biggl[
F_g(\sqrt{z}) - F_g\biggl(\frac{x_c\rho_2}{\sqrt{z}}\biggr) +
g(\frac{\sqrt{z}}{\rho_2})\biggl(g(\sqrt{z}) - g\biggl(\frac{x_c\rho_2}
{\sqrt{z}}\biggr)\biggr)\biggr] - $$
$$ - \int\limits_{x_c\rho_4}^{1}\frac{dz}{z^2}L^2\biggl[
F_g\biggl(\frac{\sqrt{z}}{\rho_4}\biggr) - F_g\biggl(\frac{x_c}
{\sqrt{z}}\biggr) + g(\sqrt{z})\biggl(g\biggl(\frac{\sqrt{z}}{\rho_4}\biggr)
- g\biggl(\frac{x_c}{\sqrt{z}}\biggr)\biggr)\biggr] - $$
\begin{equation}\label{63}
- \int\limits_{x_c\rho_3\rho_2}^{\rho_2^2}\frac{dz}{z^2}L^2\biggl[
F_g\biggl(\frac{\sqrt{z}}{\rho_3}\biggr) - F_g\biggl(\frac{x_c\rho_2}
{\sqrt{z}}\biggr) + g\biggl(\frac{\sqrt{z}}{\rho_2}\biggr)\biggl(g\biggl
(\frac{\sqrt{z}}{\rho_3}\biggr) - g\biggl(\frac{x_c\rho_2}{\sqrt{z}}\biggr)
\biggr)\biggr]\biggr\} \ .
\end{equation}
In the r.h.s. of Eq.(63) the figures into brackets are suitable for CES, when
only initial state radiation it needs to take into account.
\begin{center}
\section{ Third order correction }
\end{center}

Inside the required accuracy it needs to keep only leading contribution into 
the third
order correction. The latter becomes more important than next-to leading one
for LEP2 because of increase of the energy. In order to evalulate it one can
use the iteration up to the third order of the master equation for the electron
structure function [13]
\begin{equation}\label{64}
D(x,\alpha_{eff}) = D^{NS}(x,\alpha_{eff}) +  D^{S}(x,\alpha_{eff}) \ .
\end{equation}
The iterative form of non-singlet component of Eq.(64) reads
$$ D^{NS}(x,\alpha_{eff}) = \delta(1-x) + \sum_{k=1}^{\infty}\frac{1}{k!}
\biggl(\frac{\alpha_{eff}}{2\pi}\biggr)^kP_1(x)^{\otimes k} ,$$
\begin{equation}\label{65}
\underbrace{ P_1(x)\otimes\cdots\otimes P_1(x)}_{k} = P_1(x)^{\otimes k},
\qquad P_1(x)\otimes P_1(x) = \int\limits_{\ \ \
x}^{1}P_1(t)P_1\biggl(\frac{x}{t} \biggr)\frac{dt}{t} \ .  \end{equation}

Up to third order singlet component of Eq.(64) looks as follows [13]
\begin{equation}\label{66}
D^{S}(x,\alpha_{eff}) = \frac{1}{2!}\biggl(\frac{\alpha_{eff}}{2\pi}\biggr)^2
R(x) + \frac{1}{3!}\biggl(\frac{\alpha_{eff}}{2\pi}\biggr)^3\biggl[2P_1\otimes
R(x) - \frac{2}{3}R(x)\biggr] \ ,
\end{equation}
where R(x) is defined by Eq.(31). Effective coupling $\alpha_{eff}$ in Eqs.
(64) - (66) represents integral of running QED constant
\begin{equation}\label{67}
\frac{\alpha_{eff}}{2\pi} = \int\limits_{0}^{L}\frac{\alpha dt}{2\pi
(1-{\alpha t}/{3\pi})} = \frac{3}{2}\ln\biggl(1-\frac{\alpha L}
{3\pi}\biggr)^{-1}.
\end{equation}

The nonsinglet structure function describes the photon emission and pair 
production
without taking into account the identity of final fermions, while singlet one
is responsible just on identity effects.

Up to third order the electron structure function has the following form
$$ D(x,L) = \delta(1-x) + \frac{\alpha L}{2\pi}P_1(x) + \frac{1}{2}\biggl(
\frac{\alpha L}{2\pi}\biggr)^2\biggl(P_2(x) + \frac{2}{3}P_1(x) + R(x)\biggr) + $$
\begin{equation}\label{68}
\frac{1}{3}\biggl(\frac{\alpha L}{2\pi}\biggr)^3\biggl[\frac{1}{2}P_3(x) +
P_2(x) + \frac{4}{9}P_1(x) + \frac{2}{3}R(x) + R^{^p}(x)\biggr] \ , \quad
R^{^{^p}}(x) = P_1\oplus R(x) \ .
\end{equation}
For functions $P_3(x)$ and $R^{^p}(x) $ see [6,13 MS].

The factorization form of the differential cross-section [16] leads to
\begin{equation}\label{69}
\Sigma^{L} = \int\limits_{0}^{\infty}\frac{dz}{z^2}\int\limits_{
x_c}^{1}dx_1\int\limits_{{x_c}/{x_1}}^{1}dx_2C(x_1,L)C(x_2,L) \ ,
\end{equation}
$$C(x_1,L) = \int\limits_{x_1}^{1}\frac{dt}{t}D(t)D\biggl(\frac{x_1}{t}\biggl)
\Delta_{31}^{(t)}\ ,\ \ 
C(x_2,L) = \int\limits_{x_2}^{1}\frac{dt}{t}D(t)D\biggl(\frac{x_2}{t}\biggl)
\Delta_{42}^{(t)} \ .$$

The expansion of $C(x_1,L)$ reads
$$ C(x_1,L) = \delta(1-x_1)\Delta_{31}^{(x_1)} + \frac{\alpha L}{2\pi}P_1(x_1)
(\Delta_{31}^{(x_1)} + \Delta_{31}) + $$
$$+ \biggl(\frac{\alpha L}{2\pi}\biggr)^2\biggl[C_2(x_1)(\Delta_{31}^{(x_1)} +
\Delta_{31}) + \int\limits_{x_1}^{1}\frac{dt}{t}\Delta_{31}^{(t)}
\overline C_2(x_1,t)\biggr] + $$
\begin{equation}\label{70}
+ \biggl(\frac{\alpha L}{2\pi}\biggr)^3\biggl[C_3(x_1)(\Delta_{31}^{(x_1)} +
\Delta_{31}) + \int\limits_{x_1}^{1}\frac{dt}{t}\Delta_{31}^{(t)}
\overline C_3(x_1,t)\biggr] \ ,
\end{equation}
$$ C_2(x) = \frac{1}{2}P_2(x) + \frac{1}{3}P_1(x) + \frac{1}{2}R(x), \qquad
\overline C_2(x,t) = P_1(t)P_1\biggl(\frac{x}{t}\biggr) \ , $$
$$ C_3(x) = \frac{1}{6}P_3(x) + \frac{1}{3}P_2(x) + \frac{4}{27}P_1(x) +
\frac {2}{9}R(x) + \frac{1}{3}R^{^p}(x)\ , $$ \begin{equation}\label{71}
\overline C_3(x,t) = P_1(t)C_2\biggl(\frac{x}{t}\biggr) + C_2(t)P_1\biggl(\frac{x}
{t}\biggr)\ ,
\end{equation}
and the same for $C(x_2,L)$ with the substitution $x_2$ instead of $x_1$ and
$\Delta_{42}^{(x_2)}\ (\Delta_{42})$ instead of $\Delta_{31}^{(x_1)}\
(\Delta_{31})\ .$

Because of $\theta$ -functions under integral sign one has to distinguish
between
$$\int\limits_{x}^{1}\frac{dt}{t}A(t)B\left(\frac{x}{t}\right)\Delta_{31}^{(t)}
\quad\mbox{and}\quad \int\limits_{x}^{1}\frac{dt}{t}B(t)A\left(\frac{x}{t}
\right)\Delta_{31}^{(t)}\ .$$

In the case of CES one has to acount the initial-state radiation only.
Therefore instead of (70) it needs to write
\begin{equation}\label{72}
C_{CES}(x_1,L) = \Delta_{31}^{(x_1)}\biggl[\delta(1-x_1) +  \frac{\alpha L}
{2\pi}P_1(x_1) + \biggl(\frac{\alpha L}{2\pi}\biggr)^2C_2(x_1) + \biggl(
\frac{\alpha L}{2\pi}\biggr)^3C_3(x_1)\biggr]\ ,
\end{equation}
and analogous for $C(x_2,L).$

The last step is to write third order contribution in r.h.s. of Eq.(69):
\begin{equation}\label{73}
\Sigma_3^L = \biggl(\frac{\alpha}{2\pi}\biggr)^3\int\limits_{0}^{\infty}
\frac{dz}{z^2}L^3\int\limits_{x_c}^{1}dx\biggl(Z_1 + \int\limits_{
{x_c}/{x}}^{1}dx_1Z_2\biggr)\ ,
\end{equation}
$$ Z_1 = (2\Delta_{42}+\Delta_{42}^{(x)}\Delta_{31}+\Delta_{31}^{(x)}\Delta_
{42})C_3(x) + \int\limits_{x}^{1} \frac{dt}{t}(\Delta_{42}
^{(t)}\Delta_{31} + \Delta_{31}^{(t)}\Delta_{42})\overline C_3(x,t)\ ,$$
$$ Z_2 = [(\Delta_{31}+\Delta_{31}^{(x)})(\Delta_{42}+\Delta_{42}^{(x_1)}) +
(\Delta_{31}+\Delta_{31}^{(x_1)})(\Delta_{42}+\Delta_{42}^{(x)})]P_1(x)C_2(x_1)
+ $$
$$+ P_1(x)\int\limits_{\ \ \ \ x_1}^{1}[\Delta_{31}^{(t)}\Delta_{42} +
\Delta_{42}^{(t)}\Delta_{31} + \Delta_{31}^{(x)}\Delta_{42}^{(t)} +
\Delta_{42}^{(x)}\Delta_{31}^{(t)}]\frac{dt}{t}\overline C_2(x_1,t)\ .$$
When writing expressions for $Z_1$ and $Z_2$ it is taken into account that
$\Delta_{31}\Delta_{42} = \Delta_{42} .$ In the case of CES the expressions
for $Z_1$ and $Z_2$ may be written as follows:
\begin{equation}\label{74}
Z_1 = (\Delta_{42}^{(x)}\Delta_{31}+\Delta_{31}^{(x)}\Delta_{42})C_3(x)\ ,
\quad Z_2 = (\Delta_{42}^{(x)}\Delta_{31}^{(x_1)} + \Delta_{42}^{(x_1)}
\Delta_{31}^{(x)})P_1(x)C_2(x_1)\ .
\end{equation}

Using the relations given in Appendix B the r.h.s. of Eq.(73) may be
represented in the form suitable for numerical calculations as double integral
relative $z$- and $x$-variables. It may be written as follows:
\begin{equation}\label{75}
\Sigma_3^L = \Sigma_3^0 + \Sigma_0^3 + \Sigma_2^1 + \Sigma_1^2 \ ,
\end{equation}
where upper (down) index shows the number of additional particles (real and
virtual) emitted by the electron (the positron). The one-side emission 
contribute to the r.h.s. of Eq.(75) as
$$\Sigma_3^0 + \Sigma_0^3 = \left(\frac{\alpha}{2\pi}\right)^3\biggl\{\int
\limits_{\rho_2^{^2}}^{\rho_4^{^2}}\frac{dz}{z^2}L^{^3}\biggl[-2\int
\limits_{0}^{x_c}F_p(x)dx + 2\int\limits_{x_c}^{1}F_r(x)dx - $$
$$- \overline\theta_3^{(x_c)}\int\limits_{x_c}^{\sqrt{z}/\rho_3}F_{pr}
(x,x_c)dx\biggr] -  \int\limits_{1}^{\rho_4^{^2}}\frac{dz}{z^2}
L^{^3}\overline\theta_4^{(x_c)}\int\limits_{x_c}^{\sqrt{z}/\rho_4}
F_{pr}(x,x_c)dx +$$
\begin{equation}\label{76}
+ \int\limits_{1}^{\rho_2^{^2}}\frac{dz}{z^2}
L^{^3}\overline\theta_2^{(x_c)}\int\limits_{x_c}^{\sqrt{z}/\rho_2}
F_{pr}(x,x_c)dx\biggr\} \ ,
\end{equation}
where
$$F_p(x) = \frac{4}{3}P_3(x) + \frac{4}{3}P_2(x) + \frac{8}{27}P_1(x) , \ \ \
F_r(x) = \frac{4}{9}R(x) + \frac{5}{3}R^{^p}(x)\ , $$
$$F_{pr}(x,x_c) = \frac{1}{6}P_3(x) + \frac{1}{2}P_2(x)[\frac{2}{3} + g(\frac
{x_c}{x})] + P_1(x)[\frac{4}{27} + \frac{1}{2}f(\frac{x_c}{x}) + $$
$$+ \frac{2}{3}g(\frac{x_c}{x}) + \frac{1}{2}r(\frac{x_c}{x};1)] +
R(x)[\frac{2}{9} + \frac {1}{2}g(\frac{x_c}{x})] + \frac{1}{3}R^{^p}(x)\ , $$
 $$ r(z,1) = \int\limits_{z}^{1}R(x)dx = -\frac {22}{9} +
z + z^2 + \frac{4}{9}z^3 - \biggl(\frac{4}{3} + 2z +z^2\biggr)\ln z \ ,$$ 
$$f(z) = - F_2(z) \ .$$

In the case of CES the corresponding contribution may be derived by insertion
of functions $F_p^{^c},\  F_r^{^c}$ and $F_{pr}^{^c} $ into the r.h.s of 
Eq.(76) instead of functions $F_p,\  F_r $ and $F_{pr}$, respectively,  where
$$F_{pr}^{^c}(x) = C_3(x),\ \ \ F_p^{^c}(x) = \frac{1}{6}P_3(x) + \frac{1}{3}
P_2(x) + \frac{4}{27}P_1(x)\ , 
F_r^{^c}(x) = \frac{2}{9}R(x) + \frac{1}{3}R^{^p}(x)\ .$$

The contribution due to opposite-side emission to r.h.s. of Eq.(75) reads
$$\Sigma_2^1 + \Sigma_1^2 = \biggl(\frac{\alpha}{2\pi}\biggr)^3\biggl\{
\int\limits_{\rho_2^{^2}}^{\rho_4^{^2}}\frac{dz}{z^2}L^{^3}\biggl[
\int\limits_{0}^{x_c}\biggl(-8P_3(x) - \frac{8}{3}P_2(x)\biggr)dx +$$
$$+ 4\int\limits_{x_c}^{1}R^{^p}(x)dx -
\overline\theta_3^{(x_c)}\int\limits_{x_c}^{\sqrt{z}/\rho_3}\biggl(
H(x,x_c) + 2g(\frac{x_c}{x})h(x;\sqrt{z}/\rho_3)\biggr)dx\biggr] - $$
$$- \int\limits_{1}^{\rho_4^{^2}}\frac{dz}{z^2}L^{^3}\overline\theta_4^
{(x_c)}\int\limits_{x_c}^{\sqrt{z}/\rho_4}\biggl(H(x,x_c) + 2g(\frac
{x_c}{x})h(x;\sqrt{z}/\rho_4)\biggr)dx +$$
$$ + \int\limits_{1}^{\rho_2^{^2}}\frac{dz}{z^2}L^{^3}\overline\theta_2^
{(x_c)}\int\limits_{x_c}^{\sqrt{z}/\rho_2}\biggl(H(x,x_c) + 2g(\frac
{x_c}{x})h(x;\sqrt{z}/\rho_2)\biggr)dx +$$
$$+ \int\limits_{x_c\rho_3\rho_4}^{\rho_4^{^2}}\frac{dz}{z^2}L^{^3}
\biggl[\int\limits_{{x_c\rho_4}/{\sqrt{z}}}^{\sqrt{z}/\rho_3}\biggl(
P_1(x)G\biggl(\frac{x_c}{x};\frac{\sqrt{z}}{\rho_4}\biggr) + g\biggl(\frac
{x_c}{x};\frac{\sqrt{z}}{\rho_4}\biggr)h\biggl(x;\frac{\sqrt{z}}{\rho_3}\biggr)
\biggr)dx + (\rho_3 \leftrightarrow \ \rho_4)\ \biggr] +$$
$$+ \int\limits_{x_c\rho_2}^{1}\frac{dz}{z^2}L^{^3}
\biggl[\int\limits_{{x_c\rho_2}/{\sqrt{z}}}^{\sqrt{z}/1}\biggl(
P_1(x)G\biggl(\frac{x_c}{x};\frac{\sqrt{z}}{\rho_2}\biggr) + g\biggl(\frac
{x_c}{x};\frac{\sqrt{z}}{\rho_2}\biggr)h\biggl(x;\frac{\sqrt{z}}{1}\biggr)
\biggr)dx + (\rho_2 \leftrightarrow \ 1)\ \biggr] -$$
$$- \int\limits_{x_c\rho_3\rho_2}^{\rho_2^{^2}}\frac{dz}{z^2}L^{^3}
\biggl[\int\limits_{{x_c\rho_2}/{\sqrt{z}}}^{\sqrt{z}/\rho_3}\biggl(
P_1(x)G\biggl(\frac{x_c}{x};\frac{\sqrt{z}}{\rho_2}\biggr) + g\biggl(\frac
{x_c}{x};\frac{\sqrt{z}}{\rho_2}\biggr)h\biggl(x;\frac{\sqrt{z}}{\rho_3}\biggr)
\biggr)dx + (\rho_3 \leftrightarrow \ \rho_2)\ \biggr] -$$
\begin{equation}\label{77}
- \int\limits_{x_c\rho_4}^{1}\frac{dz}{z^2}L^{^3}
\biggl[\int\limits_{{x_c\rho_4}/{\sqrt{z}}}^{\sqrt{z}/1}\biggl(
P_1(x)G\biggl(\frac{x_c}{x};\frac{\sqrt{z}}{\rho_4}\biggr) + g\biggl(\frac
{x_c}{x};\frac{\sqrt{z}}{\rho_4}\biggr)h\biggl(x;\frac{\sqrt{z}}{1}\biggr)
\biggr)dx + (\rho_4 \leftrightarrow \ 1)\ \biggr] ,
\end{equation}
where
$$g(a;b) = g(a) - g(b) ,\ \ G(a;b) = G(a) - G(b) ,\ \ G(z) = \frac{1}{2}f(z)
+ \frac{1}{3}g(z) + \frac{1}{2}r(z)\ ,  $$
$$H(x,x_c) = P_1(x)[2f(\frac{x_c}{x}) + \frac{4}{3}g(\frac{x_c}{x}) + r(\frac
{x_c}{x};1)] + g(\frac{x_c}{x})[P_2(x) + R(x)] \ , $$
\vspace{1mm}
$$ h(x;\sqrt{z}/\rho) = \int\limits_{\ \ \ \ x}^{\sqrt{z}/\rho}\frac{dt}{t}P_1
(t)P_1\biggl(\frac{x}{t}\biggr) = $$
$$\frac{1+x^2}{1-x}\biggl(\frac{3}{2} + 2\ln\frac{(\sqrt{z}/\rho-x)(1-x)}{(1-
\sqrt{z}/\rho)x}\biggr) - 1 + x - \frac{\sqrt{z}}{\rho} + \frac{x\rho}{\sqrt{z}}
- (1+x)\ln\frac{\sqrt{z}}{x\rho}\ . $$
Note that substitutions inside square brackets concern both, limits
of $x$--integration and expressions under $x$--integral sign.

In the case of CES the r.h.s. of Eq.(77) requires the following modifications:
i) coefficient at $P_3(x)$ has to be reduced eight times, coefficients at
$P_2(x)$ and $R^{^p}(x)$ -- four times; $\;$ ii) it needs to suppouse $h$ = 0 and
to substitute $H^{^c}(x,x_c)$ instead of $H(x,x_c)$, where
$$ H^{^c}(x,x_c) = P_1(x)\biggl[\frac{1}{2}f(\frac{x_c}{x}) + \frac{2}{3}g(
\frac{x_c}{x}) + \frac{1}{2}r(\frac{x_c}{x};1)\biggr] + \frac{1}{2}g(\frac{x_c}
{x})[P_2(x) + R(x)]\ . $$
\begin{center}
\section{ The numerical results }
\end{center}

The numerical calculations carried out for the beam energy  $\epsilon = 46.15
 GeV, $ and limited angles of circular detectors as given after Eq.(3).
The Born cross-section 
\begin{equation}\label{78}
 \sigma_B =
\frac{4\pi\alpha^2}{Q_1^2}\int\limits_{\rho_2^2}^{\rho_4^2}
\frac{dz}{z^2}\left(1 - \frac{z\theta_1^2}{2}\right)
\end{equation}
(in symmetrical wide-wide case the limits of integration are 1 and $\rho_3^2)
$ equals 175.922nb for {\bf ww} angular acceptance and 139.971nb for
{\bf nn} and {\bf wn} ones. Formula (78) takes into account the contributions
of the scattered diagram as well as the interference of scattered and 
annihilation ones. The contribution of pure annihilation diagram is 
proportional to $\theta_1^4$ and is negligible even on the born level. Note,
that one has to reduce twice the coefficient at $\theta_1^2$ under integral
sign in the r.h.s. of Eq.(78) if he want restrict himself with the contribution
of the scattered diagram only.  
When
calculating the QED corrections to the cross--section (78) I systematically
ignore the terms proportional $\theta_1^2,$ which have the double logarithmic
asymptotic behavior [17] and equal parametrically to $~(\alpha|t|)\ln^2(|t|/s)  
/(\pi s).~$ The last value is about 0.1 $per \ mille$ as compared with unit
for LEP1 conditions.

The results of the numerical calculations of QED correction with the switched 
off vacuum polarization are shown in the {\bf Tables 1--3} . For comparsion we
give also the corresponding numbers derived by the help of Monte Carlo
program {\bf BLUMI} based on the YFS exponentiation [3].

As one can see from the {\bf Table1} there is an approximately constant
difference on the level of 0.3 $per\ \ mille$ between our analytical and MC
results inside first order correction. Because {\bf BLUMI} compute the first
order correction exactly [18] it may be think that this distiguish is caused
by omitted in the present calculation terms mentioned above. 

\vspace{0.5cm}
\begin{center}
\begin{tabular}{|ccccccccc|}  \hline
\multicolumn{9}{|c|}{\bf first order correction \hspace{3.0cm}
 second order correction}  \\ \hline
$x_c$   & {\bf blumi ww}  & {\bf ww}  & {\bf nn}  & {\bf wn} &
{\bf blumi ww}  & {\bf ww} & {\bf nn} & {\bf wn} \\ \hline
0.1 & 166.046 & 166.008 & 130.813 & 134.504 &
166.892 & 166.958 & 131.674 & 134.808 \\
0.3 & 164.740 & 164.702 & 129.797 &
133.416 & 165.374 & 165.447& 130.524 & 133.583 \\
0.5 & 162.241 & 162.203 &
128.001 & 131.428 & 162.530 & 162.574 & 128.474& 131.127 \\
0.7 & 155.431 &
155.390 & 122.922 & 125.809 & 155.668 & 155.597 & 123.206 & 125.225\\
0.9 &
134.390 & 134.334 & 106.478 & 107.945 &137.342 & 137.153 & 108.820 &
109.667\\
\hline
\end{tabular}
\end{center}
\hspace{0.5cm} {\small{\bf Table1}. {The SABS cross-section (in {\bf nb})
with first and second order photonic correction\\}} \vspace{0.5cm}

In the {\bf Table2} I give the absolute values of the second order
correction to SABS cross-section taking into account both leading and
next-to-leading contributions. The correction due to pair production
is small in accordance with the results of the work [6]. The second order
photonic correction is represented as a sum of leading contribution and
next-to-leading one. As one can see the next-to-leading part is not
negligible . 

 \vspace{0.5cm} \begin{center} \begin{tabular}{|rcrrccc|}
\hline \multicolumn{7}{|c|}{\bf pair production \hspace {3.cm} two photon
emission} \\ \hline $x_c$ & {\bf ww} & {\bf nn} & {\bf wn} & {\bf ww} & {\bf
nn} & {\bf wn} \\ \hline 0.1 & 0.007 & -- 0.004 & 0.015 & 0.742+0.208 &
0.679+0.182 & 0.249+0.091\\ 0.3 &-- 0.033 & -- 0.033 & -- 0.020 &
0.546+0.199& 0.556+0.171& 0.069+0.098 \\ 0.5 & -- 0.058 & -- 0.050 & -- 0.041
& 0.140+0.231& 0.291+0.182& -- 0.314+0.134 \\ 0.7 & -- 0.090 & -- 0.074 & --
0.069 & -- 0.027+0.234& 0.117+0.187& -- 0.571+0.170 \\ 0.9 &  -- 0.142 & --
0.115 & -- 0.115 & 2.961--0.142& 2.458--0.116& 1.822--0.090 \\ \hline
\end{tabular} \end{center} \hspace{2.cm} {\small{\bf Table2}.  {The second order
absolute correction to SABS cross-section (in {\bf nb})}} \vspace{0.5cm}

In the {\bf Table3} the absolute value of the leading third order correction 
and SABS cross-section with all corrections obtained in this work are shown. 
The third order one takes into account three photon
emission and pair production accompanied by single photon radiation. At large
values of $x_c$ this correction is comparable with second order
next-to-leading one. This effect will increase in the conditions of LEP2.

\vspace{0.5cm}
\begin{center}\begin{tabular}{|crrrccc|} \hline
\multicolumn{7}{|c|}{\bf third order correction \hspace{1.cm}
SABS cross-section at LEP1} \\
\hline
$x_c$ & {\bf ww} & {\bf nn} & {\bf wn} & {\bf ww} & {\bf nn} & {\bf wn} \\
\hline
0.1 & -- 0.055 & -- 0.047 & -- 0.006 & 166.910& 131.623& 134.817\\
0.3 & -- 0.065& -- 0.053 & -- 0.018& 165.349& 10.438& 133.545 \\ 0.5 & --
0.036 & -- 0.040 &  0.004& 162.472& 128.384& 131.090 \\ 0.7 &  0.089 &
0.058 &  0.124& 155.596& 123.190& 125.310 \\ 0.9 &  0.291 &  0.220 &
0.331& 137.307& 108.927& 109.893 \\ \hline
\end{tabular}\end{center} \hspace{2.0cm}{\small{\bf {Table3}}. {Leading
third order correction and SABS cross-section as obtained in this work\\}}

\vspace{0.5cm}
As concerns the second order correction it needs to have the analytical formulae
based on exponentiated form of the electron structure function in order to be 
consequent in the comparison with the {\bf BLUMI} results. On the other hand,
the comparison of given here the second order photonic correction, which 
includes the leading and next-to-leading contributions, with the corresponding
numbers for non-exponentiated {\bf BLUMI} version [3] was done recently in [22],
and the agreement is very impressive.  
\begin{center}
\section{ Conclusion }
\end{center}

In this paper analytical calculation of QED correction to SABS cross section
at LEP1 are given for the case of inclusive event selection and wide-narrow
angular acceptance. These include leading and next-to-leading contributions
in first and second orders of perturbation theory and leading one in the third
order. The leading contributions in the case of calorimeter event selection
are obtained too for any form of final electron and positron clusters. Results
are represented in the form of manifold integrals with definite limits, and
functions under integral sign have not any physical singularities. No problem
arises with infrared divergence and double counting.

The selection of essential Feynman diagrams, utilization of natural for this
problem Sudakov's variables, impact factor representation of differential
cross section due to t-channel photon exchange as well as electron structure
function method and investigation of underlying kinematics were very useful
along of the whole this work. It needs to emphasize separately the simplifications
connected with impact factor representation which allows to
represent the differential cross sections of two-jets processes in QED by
factorized form. The latter allows to use cut-off $\theta$ functions for the
final electron and positron independently on the level of the differential
cross-section. The calculation does not require to go to c.m.s. of underlying
subprocess (as in [6]) and escapes corresponding complications.

At this point I want to comment the analytical calculation of leading
contribution due to photon emission and pair production carried out in [6].
Authors of these articles
used as the master formula for description QED corrections to the SABS
cross-section due to initial-state radiation the
representation valid for cross sections of Drell-Yan  process [19],
electron-positron annihilation into muons (or hadrons) [20] and large angle
Bhabha scattering [21]. But inside this set the SABS process has a very
particular feature caused by the existence of two different scales. The
first one is the momentum transfer squared $t$, and just this scale defines the
value of the cross-section. The second scale is full c.m.s. energy squared $s =
4\epsilon^2$, and the quantity $\theta^2 \sim |t|/s << 1$ has status of a 
small correction.

The $t$-scale physics is very simple and defined by peripheral interaction
of the electron and the positron due to one photon exchange, provided momentum
transfer is pure perpendicular : $t = - \vec q^2$. The $s$-scale physics is
more complicated. On the born level it exhibits as contribution of an annihilation
diagram and beside this permits the energy and longitudinal momentum exchange
for the contribution of scattering diagram. The first order QED correction for
$s$-scale cross-section includes the contributions of box diagrams, large angle 
photon emission and up--down interference because both, the eikonal 
representation for the scattering amplitude and the factorization form of the
differential cross--section, breaks down. In the
second order large angle pair production and  appear.

The structure function used in [6] controls $t$-scale cross-section only and
has not any relation to $s$-scale one because physics of different scales
evolute by its own laws. 

On the other hand, only scattered diagram contributes in born cross-section
used in [6]. But everytime when somebody neglects annihilation diagram as
compared with scattering one he must automatically neglect $\theta^2$ as
compared with unit everywhere including the born cross-section (see comments to 
Eq.(78)) and experimental
cuts in order to be consequent. Taking into account these arguments the master
formula in [6] must be necessary simplified by eliminating terms proportional
$\xi \sim |t|/s << 1$ and $\xi^2$ in the numerator of Eq.(5) and in the cutoff
restrictions. After this it becames adequate to one obtained in [10] and
used in this work.

Numerical evaluations shows good agreement with Monte Carlo
calculations inside first order but the achievement of an
agreement for high order corrections will require an additional efforts,
connected with writing the version, based on the exponentiated form of the
electron structure function for present analytical calculation. 

\begin{center}
{\large\bf Acknowlegement \\}
\end{center}

Author thanks E. Kuraev, L. Trentadue, S. Jadach, B.W.L. Ward, G. Montagna and
B. Pietrzyk  for fruitful discussions 
and critical remarks as well as A. Arbuzov and G. Gach for the help in the
numerical calculation. This work supported by INTAS grant 93-1867.
\section*{Appendix~A}
%\begin{center} {\large\bf Appendix A \\} \end{center}
\setcounter{equation}{0} \renewcommand{\theequation}{A.\arabic{equation}}

Let us begin with the consideration of the next-to-leading second order
$\Delta$-independent contribution due to one-side two photons emission. At
first I will give analytical expression for symmetrical case, because it was
not published up to now in relevant form. (I do not introduce special notation for 
next-to-leading contribution to $\Sigma$ keeping in mind that only such kind 
terms are considered along this Appendix)
\begin{equation} \label{ea1}
\Sigma^{\gamma\gamma} = \Sigma_{\gamma\gamma} =
\frac{1}{4}\left(\frac{\alpha}{\pi}\right)^2\int\limits_{1}^{\rho^2}
\frac{dz}{z^2}L\; Y, \end{equation}
$$Y = y + \int\limits_{x_c}^{1}dx\;\biggl\{A + \int\limits_{0}^
{1-x}dx_1\; \biggl[\frac{1}{x_1}\; 4\;\frac{1+x^2}{1-x} (\theta_{\rho}^{(x)}
l_1+l_2) + \biggl(-1-\frac{1+x}{1-x_1} -$$
$$-\frac{x}{(1-x_1)^2}\biggr)(l_4 + \theta_{\rho}^{(x)}l_3 + 2\theta_{\rho}^
{(1-x_1)}l_5) + \frac{2(1+x)}{1-x_1}\theta_{\rho}^{(1-x_1)}\biggr] -$$
$$ -4\;\frac{1+x^2}{1-x}\;\overline\theta_{\rho}^{(x)} \biggl[\int\limits_
{1-\sqrt{z}/\rho}^{1-x}dx_1\; \biggl(\frac{1}{x_1}l_5 + \frac{2}{x_2}\ln
\frac{x}{1-x_1}\biggr) + \int\limits_{0}^{\sqrt{z}/\rho-x}\frac{dx_1}
{x_1}\; l_6\biggr] \biggr\}\ , $$ 
$$y = 12\zeta_3 + 10\zeta_2 - \frac{45}{4} - 16\ln^2(1-x_c) - 28\ln(1-x_c)\ ,$$
$$A = (1+\theta_{\rho}^{(x)})\biggl[2(5+2x)+4(x+3)\ln(1-x) + 4\;\frac{1+x^2}
{1-x}\ln x\biggr] +$$
$$ + 2\;\frac{1+x^2}{1-x}\biggl[(\frac{3}{2}-
\ln x)K(x,z;\rho,1) - \frac{1}{2}\ln^2x - \frac{(1-x)^2}{2(1+x^2)} +$$
$$ + 2\ln(1-x)\biggl(\theta_{\rho}^{(x)}\ln\left|\frac{x^2\rho^2-z} {x\rho^2-z}
\right|+ \ln\left|\frac{(z-1)(z-x^2)(\rho^2-z)} {(z-x)^2(x\rho^2-z)}\right|
\biggr)\biggr] +$$
$$ + \overline{\theta}_{\rho}^{(x)}\biggl[\frac{16}
{1-x}\ln(1-x) + \frac{14}{1-x} - (1-x)\ln x +$$
$$ + 2\;\frac{1+x^2}
{1-x}\biggl( - \frac{3}{2}\ln^2x + 3\ln x\ln(1-x) - L_{i2}(1-x) - \frac{x(1-x)+
4x\ln x}{2(1+x^2)} +$$
$$ + \frac{(1+x)^2}{1+x^2}\ln\left|\frac
{(\sqrt{z}-x\rho)} {\rho-\sqrt{z}}\right|+ 2\ln\left|\frac{\sqrt{z}-x\rho}
{\rho}\right| \ln\left|\frac{x(x\rho^2-z)}{x^2\rho^2-z}\right|\biggr)\biggr],$$
$$l_1 = \ln\left|\frac{(x^2\rho^2-z)(x\rho^2-z)} {(x(1-x_1)\rho^2-z)(x(x+x_1)
\rho^2-z)}\right|,\ \  l_3 = \ln\left|\frac{(1-x_1)^2(1-x-x_1)(x^2\rho^2-z)^2}
{x^3x_1(x(1-x_1)\rho^2-z)^2}\right|, $$
$$l_2 = \ln\left|\frac{(z-x)^2(z-(1-x_1)^2)(z-(x+x_1)^2)} {(z-(1-x_1))
(z-x(1-x_1))((x+x_1)-z)(x(x+x_1)-z)}\right| +$$
$$ \ln\left|\frac{((1-x_1)^2\rho^2-z)((x+x_1)^2\rho^2-z)(x\rho^2-z)} {((x+x_1)
\rho^2-z)((1-x_1)\rho^2-z)(x^2\rho^2-z)}\right|, $$
$$l_4 = \ln\left|\frac{(1-x_1)^2xx_1(z-1)(z-x^2)(z-(1-x_1)^2)^2} {x_2(z-(1-x_1))^
2(z-x(1-x_1))^2}\right| + \ln\left|\frac{(\rho^2-z)(x(1-x_1)
\rho^2-z)^2} {(x^2\rho^2-z)((1-x_1)^2\rho^2-z)^2}\right|, $$
$$l_5 = \ln\left|\frac{x((1-x_1)^2\rho^2-z)^2} {(1-x_1)^2(x(1-x_1)\rho^2-z)
((1-x_1)\rho^2-z)^2}\right|, $$
$$l_6 = \ln\left|\frac{(x\rho^2-z)((x+x_1)^2\rho^2-z)^2} {(x^2\rho^2-z)
(x(x+x_1)\rho^2-z)((x+x_1)\rho^2-z)}\right|.$$

For wide-narrow angular acceptance it needs to consider only the case of the
positron emission $\Sigma_{\gamma\gamma}$, because the corresponding expression
for the electron emission $\Sigma^{^{\gamma\gamma}}$ is just eq.(A1) with
$(\rho_4^2,\rho_2^2)$ as the limits of $z$-integration and $\rho_3$ instead of
$\rho$ under the integral sign.

The analytical expression for $\Sigma_{\gamma\gamma}$ has the following form:
\begin{equation} \label{ea2}
\Sigma_{\gamma\gamma} = \frac{1}{4}\left(\frac{\alpha}{\pi}\right)^2\int
\limits_{1}^{\rho_3^2}\frac{dz}{z^2}L\; A^{W}_{N}\, , \end{equation}
$$A^{W}_{N} = y\Delta_{42} + \int\limits_{x_c}^{1}dx\Biggl\{ \Delta_{42}
\biggl[4(4+3x)+6(x+3)\ln(1-x)+\biggl(x-1 + $$
$$+ 4\;\frac{1+x^2}{1-x}
\biggr)\ln x\biggr] + \Delta_{42}^{(x)}\biggl[(1-x)(3+\ln x) + 2(x+3)\ln(1-x)
+ 4\;\frac{1+x^2}{1-x}\ln x \biggr] + \overline{\Delta}_{42}^
{(x)} \frac{2}{1-x}(4+$$ 
$$+ (1+x)^2)\ln(1-x) + 2\;\frac{(1+x)^2}{1-x}
\biggl(\theta_4\overline{\theta}_4^{(x)} \ln\left|\frac{\sqrt{z}-x\rho_4}
{\rho_4-\sqrt{z}}\right| - \theta_2\overline{\theta}_2^{(x)}\ln\left|\frac
{\sqrt{z}-x\rho_2} {\rho_2-\sqrt{z}}\right|\biggr) + $$ 
$$+ \frac{1+x^2}{1-x}B+ \int\limits_{0}^{1-x}dx_1\biggl[2\;\frac{1+x^2}
{(1-x)x_1} \biggl(\Delta_{42}^{(x)}l_{1+} + \Delta_{42}l_{2+} +
(\overline{\theta}_4^{(x)} - \theta_2^{(x)})l_{1-} +
(\overline{\theta}_4 -\theta_2)l_{2-} \biggr) + $$ 
$$+ \biggl(-1-\frac{1+x}{1-x_1}-
\frac{x}{(1-x_1)^2}\biggr)\Biggl(\Delta_{42}^{(x)}\biggl(\ln\frac{(1-x_1)
^2x_2}{x^3x_1} + l_{3+}\biggr) + \Delta_{42}\biggl(\ln\frac{(1-x_1)^2xx_1}
{x_2} + l_{4+}\biggr) + $$ 
$$+ \Delta_{42}^{(1-x_1)}\biggl(2\ln\frac{x}{(1-x_1)^2} + l_{5+}\biggr)
+ (\overline{\theta}_4^{(x)}-\theta_2^{(x)})l_{3-} + (\overline{\theta}_4-
\theta_2)l_{4-} +$$ 
$$+ (\overline{\theta}_4^{(1-x_1)}-\theta_2^
{(1-x_1)})l_{5-}\Biggr) + 2\;\frac{1+x}{1-x_1}\Delta_{42}^{(1-x_1)} \biggr]
+ 2\;\frac{1+x^2}{1-x}\theta_4\overline{\theta}_4^{(x)}
\biggl[ \int\limits_{1-\sqrt{z}/\rho_4}^{1-x}dx_1 \biggl(\frac{1}{x_1}
\overline{l}_6 - $$ 
$$- \frac{4}{x_2}\ln\frac{x}{1-x_1}\biggr) + \int\limits_{0}
^{\sqrt{z}/\rho_4-x}\frac{dx_1}{x_1}\overline{l}_7\biggr]
+ 2\;\frac{1+x^2}{1-x}\theta_2\overline{\theta}_2^{(x)}\biggl[ \int\limits_
{1-\sqrt{z}/\rho_2}^{1-x}dx_1 \biggl(\frac{1}{x_1}\tilde{l}_6 + $$ 
$$+ \frac{4}{x_2}
\ln\frac{x}{1-x_1}\biggr) + \int\limits_{0}^{\sqrt{z}/\rho_2-x}\frac
{dx_1}{x_1}\tilde{l}_7\biggr] \Biggr\}, $$ 
$$B = \Delta_{42}\biggl(-2\ln^2x+2\ln(1-x) \ln\left|\frac{x^4(z-\rho_2^2)^2
(z-x^2\rho_2^2)(x^2\rho_4^2-z)(\rho_4^2-z)^2} {(z-x\rho_2^2)^3(x\rho_4^2-z)^3}
\right|\biggr) + $$ 
$$+ \Delta_{42}^{(x)}\biggl(\ln^2x+2\ln(1-x)
\ln\left|\frac{(z-x^2\rho_2^2)(x^2\rho_4^2-z)} {x^4(z-x\rho_2^2)(x\rho_4^2-z)}
\right|\biggr) + (3-2\ln x)\widetilde{K}(x,z;\rho_4,\rho_2) + $$ 
$$+ \overline{\Delta}_{42}^{(x)}\biggl(7-2\ln x\ln(1-x) - 2\ln^2x
- 2L_{i2}(1-x) - \frac{x(1-x)+4x\ln x}{1+x^2}\biggr) + $$ 
$$+ 2(\overline{\theta}_4-\theta_2)\ln(1-x)\ln\left|\frac{(x\rho_4^2-z)^3
(z-\rho_2^2)^2(z-x^2\rho_2^2)}{(\rho_4^2-z)^2(x^2\rho_4^2-z) (z-x\rho_2^2)^3}
\right| +$$ 
$$+ 2(\overline{\theta}_4^{(x)}-\theta_2^{(x)})\ln(1-x) \ln\left|\frac
{(z-x^2\rho_2^2)(x\rho_4^2-z)} {(x^2\rho_4^2-z)(x\rho_2^2-z)}\right| +$$ 
$$+ 4\theta_4\overline{\theta}_4^{(x)}\ln\left|\frac{x\rho_4-
\sqrt{z}} {\rho_4}\right|\ln\left|\frac{x(x\rho_4^2-z)}{x^2\rho_4^2-z}\right|
+4\theta_2\overline{\theta}_2^{(x)}\ln\left|\frac{\sqrt{z}-x\rho_2}
{\rho_2}\right|\ln\left|\frac{z-x^2\rho_2^2}{x(z-x\rho_2^2)}\right|, $$ 
$$l_{1\pm} = (1\pm \hat{c})\ln\left|\frac{(z-x^2\rho_2^2)(z-x\rho_2^2)}
{(z-x(1-x_1)\rho_2^2)(z-x(x+x_1)\rho_2^2)}\right|, $$ 
$$l_{2\pm}= (1\pm \hat{c})\Biggl[\ln\left|\frac{(z-x\rho_2^2)^3(z-(1-
x_1)^2\rho_2^2)^2(z-(x+x_1)^2\rho_2^2)^2} {(z-x^2\rho_2^2)(z-x(1-x_1)\rho_2^2)
(z-x(x+x_1)\rho_2^2)(z-(1-x_1)\rho_2^2)^2(z-(x+x_1)\rho_2^2)^2}\right|
\Biggr], $$ 
$$l_{3\pm} = (1\pm \hat{c})\ln\left|\frac{z-x^2\rho_2^2} {z-x(1-x_1)\rho_2^2}
\right|, \quad
l_{4\pm} = (1\pm \hat{c})\ln\left|\frac{z-\rho_2^2} {z-(1-x_1)\rho_2^2}
\right|, $$ 
$$l_{5\pm} = (1\pm \hat{c})\ln\left|\frac{(z-(1-x_1)^2
\rho_2^2)^2} {(z-x(1-x_1)\rho_2^2)(z-(1-x_1)\rho_2^2)}\right|, $$ 
$$\tilde{l}_{6} = \ln\left|\frac{x^2(z-(1-x_1)^2\rho_2^2)^4} {(1-x_1)^4
(z-x(1-x_1)\rho_2^2)^2(z-(1-x_1)\rho_2^2)^2}\right|, $$ 
$$\tilde{l}_{7} = \ln\left|\frac{(z-x\rho_2^2)^2(z-(x+x_1)^2\rho_2^2)^4}
{(z-x^2\rho_2^2)^2(z-x(x+x_1)\rho_2^2)^2(z-(x+x_1)\rho_2^2)^2}\right|\ ,\ \
\overline{l}_6 = - \hat{c}\tilde{l}_6\ , \ \overline{l}_7 = - \hat{c}
\tilde{l}_7 \ ,$$
where $x_2=1-x-x_1$, and $\hat{c}$ is the operator of the substitution
\begin{eqnarray}
\hat{c}f(\rho_2)=f(\rho_4) \ .
\end{eqnarray}
One can verify that in the symmetrical limit formula (\ref{ea2}) coincides
with (\ref{ea1}) one.

For opposite-side emission the next-to-leading contribution to $\Sigma$
in the symmetrical case reads
\begin{eqnarray} \label{ea4}
\Sigma^{\gamma}_{\gamma} = \left(\frac{\alpha}{\pi}\right)^2 L
\int\limits_{0}^{\infty}\frac{dz}{z^2}\; T, \end{eqnarray}
\begin{eqnarray}
T &=& A\theta_{\rho}\overline\theta_1 - \int\limits_{x_c}^{1}dx\;
\biggl[\frac{1+x^2}{2(1-x)}N(x,z;\rho,1) + \Xi(x) + \frac{\overline{\Xi}(x)}
{1-x}\biggr] \\ \nonumber &\times& \int\limits_{x_c/x_1}^{1}dx_1\;
\biggl[(1+x_1)\Xi(x_1) + \frac{2\overline{\Xi}(x_1)}{1-x_1}\biggr], \end{eqnarray}
where
\begin{eqnarray}
A &=& - 6 - 14\ln(1-x_c) - 8\ln^2(1-x_c) + \int\limits_{\ \ \ \ x_c}^{1}dx
\biggl\{ 7(1+x) + \\ \nonumber &+& \frac{1+x^2}{2(1-x)}[3K(x,z;\rho,1) +
7\overline{\theta}_{\rho}^{(x)}]
+ 2\ln\frac{x-x_c}{x}\biggl[(3+x)(1+\theta_{\rho}^{(x)}) + \\ \nonumber
&+& \frac{4}{1-x}\overline{\theta}_{\rho}^{(x)} + \frac{1+x^2}{1-x}
N(x,z;\rho,1)\biggr] + \frac{8}{1-x}\ln\frac{x(1-x_c)}{x-x_c}\biggr\} \ .
\end{eqnarray}
We introduce the following reduced notation for $\theta$-functions:
$$\Xi(x)=\theta_{\rho}\overline{\theta}_1+\theta_{\rho}^{(x)}
\overline{\theta}_1^{(x)}, \quad \overline{\Xi}(x)=\theta_{\rho}
\overline{\theta}_{\rho}^{(x)} - \theta_1\overline{\theta}_1^{(x)} \ $$.
The quantity $K(x,z;\rho,1)$ entering into espression for $A$ is the $K$--
factor for single photon emission, and the quantity $N(x,z;\rho,1)$ may be
derived by the help of Eq.(10) in the following way:
$$N(x,z;\rho,1)=\biggl(\widetilde{K}(x,z;\rho_4,\rho_2) - \frac{(1-x)^2}
{1+x^2}(\Delta_{42} + \Delta_{42}^{(x)})\biggr) \bigg|_{\rho_4=\rho,
\ \rho_2=1} \ .$$
Note that $N(1,z;\rho,1)=0 \ .$

In the wide-narrow angular acceptance the corresponding formula for
$\Sigma_{\gamma}^{\gamma}$ may be written as follows:
\begin{equation}
\Sigma_{\gamma}^{\gamma}=\frac{\alpha^2}{\pi^2}L \int\limits_{0}^{\infty}
\frac{dz}{z^2}T_{N}^{W}, \end{equation}
where
\begin{eqnarray}
T_{N}^{W} &=& \widetilde{A} - \frac{1}{2}\;\Biggl\{ \int\limits_{x_c}
^{1}dx\;\biggl[
\frac{1+x^2}{2(1-x)}N(x,z;\rho_3,1) + \Xi_{31}(x) + \frac{1}{1-x}\overline{
\Delta}_{31}^{(x)}\biggr] \\ \nonumber &\times& \int\limits_{x_c/x}^{1}dx_1\; 
\biggl[(1+x_1)\Xi_{42}(x) + \frac{2}{1-x_1}\overline{\Delta}_{42}^{(x)}\biggr] 
+ \\ \nonumber
&+& \int\limits_{x_c}^{1}dx\;\biggl[ \frac{1+x^2}{2(1-x)}
N(x,z;\rho_4,\rho_2) + \Xi_{42}(x) + \frac{1}{1-x}\overline{\Delta}_{42}^{(x)}
\biggr] \\ \nonumber &\times& \int\limits_{x_c/x}^{1}dx_1\;\biggl
[ (1+x_1)\Xi_{31}(x) + \frac{2}{1-x_1}\overline{\Delta}_{31}^{(x)}\biggr]
\Biggr\} \ , \end{eqnarray}
where
\begin{eqnarray}
\widetilde{A} &=& ( - 6 - 14\ln(1-x_c) - 8\ln^2(1-x_c))\Delta_{42} + \\
\nonumber &+& \int\limits_{x_c}^{1}dx\Biggl\{
\Delta_{42}\biggl[7(1+x) + \frac{8}{1-x}\ln\frac{x(1-x_c)}{x-x_c}\biggr] + \\
\nonumber
&+& \frac{1+x^2}{2(1-x)}\biggl[\frac{3}{2}\Delta_{42} \widetilde{K}(x,z;
\rho_3,1)
+ \frac{3}{2}\Delta_{31}\widetilde{K}(x,z;\rho_4,\rho_2) + \\ \nonumber
&+& \frac{7}{2}(\Delta_{42}\overline{\Delta}_{31}^{(x)} + \Delta_{31}
\overline{\Delta}_{42}^{(x)})\biggr] + \ln\frac{x-x_c}{x}\biggl[(3+x)
(\Delta_{31}\Xi_{42}(x) + \Delta_{42}\Xi_{31}(x)) + \\ \nonumber
&+& \frac{4}{1-x}(\overline{\Delta}_{42}^{(x)}\Delta_{31} + \overline{\Delta}_
{31}^{(x)}\Delta_{42}) + \frac{1+x^2}{1-x}(\Delta_{42}N(x,z;\rho_3,1) + \\
\nonumber &+& \Delta_{31}N(x,z;\rho_4,\rho_2))\biggr]\Biggr\}, \end{eqnarray}
and
$$\Xi_{42}(x) = \theta_4\overline{\theta}_2+\theta_4^{(x)}\overline{\theta}
_2^{(x)} = \Delta_{42}+\Delta_{42}^{(x)}\ , $$
$$\Xi_{31}(x)=\Delta_{31}+\Delta_{31}^{(x)}, \quad \overline{\Delta}_{31}^
{(x)}=\Delta_{31}-\Delta_{31}^{(x)}\ .$$
It is obvious that in symmetrical limit formula (A.7) coinsides with (A.4)
one.

\section*{Appendix~B}
\setcounter{equation}{0}
\renewcommand{\theequation}{B.\arabic{equation}}

Here I give some relations which were used in the process of analytical
calculations  and at numerical computations.
For the case of emission along the electron momentum direction they reads
$$\int\limits_{\rho_2^2}^{\rho_4^2}dz\; \int\limits_{x_c}^
{1}dx\;\overline{\theta}_3^{(x)} = \int\limits_{\rho_2^2}^{\rho_4^2}dz
\;\overline{\theta}_3^{(x_c)} \int\limits_{x_c}^{\sqrt{z}/\rho_3}dx \ , $$
\begin{equation}
\int\limits_{\rho_2^2}^{\rho_4^2}dz\; \int\limits_
{x_c}^{1}dx\;\int\limits_{0}^{1-x}dx_1 \overline{\theta}_3^
{(1-x_1)}  =\int\limits_{\rho_2^2}^{\rho_4^2}dz\;
\overline{\theta}_3^{(x_c)} \int\limits_{x_c}^{\sqrt{z}/\rho_3}dx\;
\int\limits_{1-\sqrt{z}/\rho_3}^{1-x}dx_1\, .
\end{equation}

For the case of the emission along the positron momentum direction: 
\begin{eqnarray}
&& \int\limits_{1}^{\rho_3^2}dz\;\int\limits_{x_c}^{1}dx\;
[\overline{\theta}_4^{(x)} - \theta_2^{(x)}] = \int\limits_{1}^
{\rho_3^2}dz\;\int\limits_{x_c}^{1}dx\; [\overline{\theta}_4-\theta_2
+\theta_4\overline{\theta}_4^{(x)} + \theta_2\overline{\theta}_2^{(x)}] 
\nonumber
\\ && \quad = \int\limits_{1}^{\rho_3^2}dz\;\biggl\{ (\overline{\theta}_
4-\theta_2)\int\limits_{x_c}^{1}dx\; + \theta_4\overline{\theta}_4^{(x_c)}
\int\limits_{x_c}^{\sqrt{z}/\rho_4}dx\; + \theta_2\overline{\theta}_2
^{(x_c)}\int\limits_{x_c}^{\sqrt{z}/\rho_2}dx \biggr\}, \\ \nonumber
&& \int\limits_{1}^{\rho_3^2}dz\;
\int\limits_{x_c}^{1}dx\;\int\limits_{0}^{1-x}dx_1\;
[\overline{\theta}_4^{(1-x_1)}-\theta_2^{(1-x_1)}] = \int\limits_{1}^
{\rho_3^2}dz\;\int\limits_{x_c}^{1}dx\;\biggl\{ (\overline{\theta}_4-
\theta_2) + \int\limits_{0}^{1-x}dx_1\; \\ \nonumber && \quad
+ \overline{\theta}_4^{(x_c)}\theta_4\int\limits_{x_c}^{\sqrt{z}/\rho_
4}dx \int\limits_{1-{\sqrt{z}}/{\rho_4}}^{1-x}dx_1+\overline{\theta}_2^
{(x_c)} \theta_2\int\limits_{x_c}^{\sqrt{z}/\rho_2}dx \int\limits_
{1-\sqrt{z}/\rho_2}^{1-x}dx_1\biggr\}.
\end{eqnarray}
Some additional relations arise for the case of the opposite-side emission. Let
us consider first the integration limits restrictions for the product of
$\theta$-functions in the symmetrical case:
\begin{equation}
\theta_3\overline{\theta}_3^{(x_1)}\overline{\theta}_3^{(x_2)}, \quad \theta_
1\overline{\theta}_3^{(x_1)}\overline{\theta}_1^{(x_2)}, \quad \theta_1
\overline{\theta}_1^{(x_1)}\overline{\theta}_1^{(x_2)}.
\end{equation}
At first it needs to use the formulae (B.1) and get rid $\;\overline{\theta}_i^{(x_2)}$
using the following changes: i)$\;\overline{\theta}_i^{(x_2)} \to
\overline{\theta}_i^{({x_c}/{x_1})}$, ~ii)$\;$ the upper limit of $x_2$
integration in the case of $\;\overline{\theta}_3^{(x_2)}$ ~has to be
replaced by $({\sqrt{z}}/{\rho_3})$ ~and in the case of
$\;\overline{\theta}_1^{(x_2)}$ by $\sqrt{z}$.

Thus, there are three regions defined by following curves in $(z,x_1)$ plane:
\begin{eqnarray}
&& \rho^2=z,\quad z=x_1^2\rho^2,\quad z=\frac{x_c^2\rho^2}{x_1^2}\, , 
\\ \nonumber && 1=z,\quad z=x_1^2\rho^2,\quad z=\frac{x_1}{x_c^2}\, , 
\\ \nonumber && 1=z,\quad z=x_1^2,\quad z=\frac{x_1^2\rho^2}{x_c^2}.
\end{eqnarray}
It easy to see that the limits of integrations may be transformed as follows: 
\begin{eqnarray}
&& \int\theta_3\overline{\theta}_3^{(x_1)}\overline{\theta}_3^{(x_2)} \
\rightarrow \int\limits_{x_c\rho^2}^{\rho^2}dz\; \int\limits_{
x_c\rho/\sqrt{z}}^{\sqrt{z}/\rho}dx_1\; \int\limits_{x_c/x_1}^
{\sqrt{z}/\rho}dx_2\; , \\ \nonumber && \int\theta_3\overline{\theta}_1^
{(x_1)}\overline{\theta}_1^{(x_2)} \rightarrow \int\limits_{x_c\rho}^
{1}dz\; \int\limits_{x_c/\sqrt{z}}^{\sqrt{z}/\rho}dx_1\; \int\limits_
{x_c/x_1}^{\sqrt{z}}dx_2\; , \end{eqnarray}
and for $\int\theta_1\overline{\theta}_1^{(x_1)}\overline{\theta}_1^{(x_2)}$
the formulae may be obtained from the above ones by putting $\rho=1$.
For the wide-narrow angular acceptance the prescription is similar:
\begin{equation}
\int\theta_4\overline{\theta}_4^{(x_1)}\overline{\theta}_3^{(x_2)}
\rightarrow \int\limits_{x_c\rho_3}^{\rho_4^2}dz\;
\int\limits_{x_c\rho_3/\sqrt{z}}^{\sqrt{z}/\rho_4}dx_1\;
\int\limits_{x_c/x_1}^{\sqrt{z}/\rho_3}dx_2\, .
\end{equation}
The another variants of restrictions in wide-narrow ansular acceptancee may
be transformed as follows:
\begin{eqnarray}
&& \int\theta_1\overline{\theta}_2^{(x_1)}\overline{\theta}_1^{(x_2)}
\rightarrow \int\limits_{x_c\rho_2}^{1}dz\; \int\limits_{
{x_c\rho_2}/{\sqrt{z}}}^{\sqrt{z}}dx_1\; \int\limits_{{x_c}/{x_1}}^{\sqrt{z}/
\rho_2}dx_2\, , \\ \nonumber && \int\theta_1\overline{\theta}_4^{(x_1)}
\overline{\theta}_1^{(x_2)} \rightarrow \int\limits_{x_c\rho_4}^{1}dz
\; \int\limits_{x_c\rho_4/\sqrt{z}}^{\sqrt{z}}dx_1\; \int\limits_{x_c/x_1}
^{\sqrt{z}/\rho_4}dx_2\, , \\ \nonumber && \int\theta_2\overline{\theta}_2^
{(x_1)}\overline{\theta}_3^{(x_2)} \rightarrow \int\limits_{\ \ x_c\rho_2
\rho_3}^{\rho_2^2}dz\; \int\limits_{x_c\rho_3/\sqrt{z}}^{\sqrt{z}/
\rho_2}dx_1\; \int\limits_{x_c/x_1}^{\sqrt{z}/\rho_3}dx_2\, .
\end{eqnarray}
\section*{References}
\begin{enumerate}
\item The LEP Collaboration: ALEPH, DELPI, L3 and OPAL and the LEP Electroweak
Working Group, CERN--PPE/95; B. Pietrzyk, preprint LAPP-Exp-94.18. Invited
talk at the Conference on "Radiative Corrections: Status and Outlook".
Galtinburg, TN, USA, 1994; I.C. Brock et al., Preprint CERN-PPE/96-89'
CMU-HEP/96-04, 27 June 1996.
\item Neutrino Counting in Z Physics at LEP,
G. Barbiellini et al., L. Trentadue (conv.); G. Altarelli, R. Kleiss and
C. Verzegnassi eds. CERN Report 89-08.
\item Events Generator for
Bhabha Scattering, H. Anlauf et al., Conveners: S. Jadach and O. Nicrosini.
Yellow Report CERN 96-01, v.2 p.229 - 298. \item S. Jadach, E. Richter-Was,
B.F.L. Ward and Z. Was, Comput.Phys.Commun. {\bf 70} (1992) 305.  
\item G. Montagna
et al., Comput.Phys.Commun. {\bf 76} (1993) 328; M. Cacciori, G. Montagna 
F. Piccinini,
Comput.Phys.Commun. {\bf 90} (1995) 301, CERN-TH/95-169; G. Montagna et al.,
Nucl.Phys.{\bf B 401} (1993) 3.  \item S. Jadach, M. Skrzypek and B.F.L. Ward,
Phys.Rev. {\bf D 47} (1993) 3733; S. Jadach, E. Richter-Was, B.F.L. Ward and 
Z. Was,
Phys.Lett. {B 260} (1991) 438.  \item S. Jadach, E. Richter-Was, B.F.L. Ward 
and
Z. Was, Phys.Lett. {B 353} (1995) 349,\ 362; \ S. Jadach, M. Melles, B.F.L. 
Ward and S.A. Yost, Phys.Lett.{\bf B 377} (1996) 168.   
\item W. Beenakker, F.A. Berends and
S.C. van der Marck, Nucl.Phys. {\bf B 355} (1991) 281;\  W. Beenakker and
B. Pietrzyk, Phys.Lett. {\bf B 304} (1993) 366.  
\item M. Gaffo, H. Czyz, E. Remiddi, Nuovo Cim. {\bf 105 A} (1992) 271;
Int.J.Mod.Phys. {\bf 4} (1993) 591;\ 
Phys.Lett. {/bf B 327} (1994) 369;\ G. Montagna, O. Nicrosini and F. Piccinini,
Preprint FNT/T--96/8.
\item A.B. Arbuzov et al., Yellow Report CERN 95-03,
p.369; \ preprint CERN-TH/95-313, UPRF-95-438 (to be published in Nucl.Phys.).
\item A.B. Arbuzov, E. Kuraev, N.P. Merenkov and L. Trentadue, JETP {\bf 81} 
(1995) 638;\
Preprint CERN-TH/95-241,\ JINR-E2-95-110.
\item D.R. Yennie, S.C. Frautchi and H. Suura, Ann.Phys. {\bf 13} (1961) 379.
\item L.N.Lipatov, Sov.J.Nucl.Phys. {\bf 20} (1974) 94;\ G. Altarelli 
and G. Parisi, Nucl.Phys.
{\bf B 126} (1977) 298;\  M. Skrzypek, Acta Phys.Pol. {\bf B 23} (1992) 135.
\item N.P. Merenkov, Sov.J.Nucl.Phys. {\bf 48} (1988) 1073; {\bf 50} (1989) 469.
\item T.D. Lee and M. Nauenberg, Phys.Rev. {\bf B 133} (1964) 1549.
\item H. Cheng and T.T. Wu, Phys.Rev.Lett. {\bf 23} (1969) 670;\ V.G. Zima 
and N.P. Merenkov,
Yad.Fis. {\bf 25} (1976) 998;\ V.N. Baier, V.S. Fadin, V. Khoze and E. Kuraev,
 Phys.Rep. {\bf 78} (1981) 294.
\item V.G. Gorshkov, Uspechi Fiz. Nauk {\bf 110} (1973) 45;\ F.A. Berends 
et al., Nucl. Phys. {\bf B 57} (1973) 371; \ E.A. Kuraev and G.V. Meledin,
Nucl. Phys.{\bf B 122} (1977) 3582. 
\item S. Jadach and B.W.L. Ward, Phys. Rev. {\bf D 40} (1989) 3582.
\item S. Drell and T.M. Yan, Phys. Rev. Lett, {\bf 25} (1970) 316.
\item E.A. Kuraev and V.S. Fadin, Sov. J. Nucl. Phys. {\bf 41} (1985) 466.
\item W. Beenakker, F.A. Berends and S.C. van der Marck, Nucl. Phys. {\bf B 
349} (1991) 323.
%\item A.P.Bukhvostov, E.A.Kuraev, L.N.Lipatov. Sov.Yad.Fis. 38 (1983) 439;
%39 (1984) 194.
%\item Solitons and Nonlinear Wave Equations. R.K.Dodd, J.C.Eilbeck, J.D.Gibbon,
%H.C.Morris. Academic Press, Inc. (Harcourt Brace Jovanovich, Publishers);
%S.S.Moiseyev, P.B.Rutkevich, A.V.Tur, V.V.Yanovsky. Sov.JETP, 94 (1988) 144;
%A.V.Chechkin, A.V.Tur, V.V.Yanovsky. Phys.Fluid B, 4 (1992) 3513.
\item A.B. Arbuzov et al., Phys. Lett. {\bf B} (in press).
\end{enumerate}
\end{document}